\documentclass[prx, twocolumn]{revtex4-2}
\usepackage[x11names, rgb, html]{xcolor}

\definecolor{red}{HTML}{f54b1a}
\definecolor{pink}{HTML}{d19eb1}
\definecolor{orange}{HTML}{d3772e}
\definecolor{yellow}{HTML}{ebe85d}
\definecolor{green}{HTML}{0f6852}
\definecolor{lightblue}{HTML}{01abe9}
\definecolor{darkblue}{HTML}{1b346c}
\definecolor{tan}{HTML}{e5c39e}
\definecolor{darktan}{HTML}{af9e73}
\definecolor{grey}{HTML}{c3ced0}
\definecolor{darkgrey}{HTML}{9dadc4}
\definecolor{black}{HTML}{110d1b}
\definecolor{white}{HTML}{f1f8f1}

\usepackage{hyperref}
\hypersetup{
  colorlinks = true,
  linkcolor = red,
  citecolor = darkblue,
  urlcolor = lightblue
}

\usepackage[cmintegrals,cmbraces]{newtxmath}

\usepackage{graphicx}
\usepackage{dcolumn}
\usepackage{bm}
\usepackage{float}
\usepackage{caption}[justification=justified]
\usepackage{subcaption}

\graphicspath{{Figures/}}

\usepackage[margin=1in]{geometry}

\usepackage{algorithm}
\usepackage[noend]{algpseudocode}
\algrenewcommand{\algorithmiccomment}[1]{$\vartriangleright$ #1}
\algrenewcommand{\algorithmicreturn}{\textbf{Return: }}
\algnewcommand\algorithmicinput{\textbf{Input: }}
\algnewcommand\Input{\State \algorithmicinput}

\newcommand{\diff}[2]{\frac{\partial#1}{\partial #2}}

\def\nb{\boldsymbol{n}}

\def\Lambdab{\boldsymbol{\Lambda}}
\def\Gammab{\boldsymbol{\Gamma}}

\def\bb{\boldsymbol{b}}

\def\eb{\boldsymbol{e}}
\def\fb{\boldsymbol{f}}
\def\qb{\boldsymbol{q}}

\def\Jb{\boldsymbol{J}}
\def\kb{\boldsymbol{k}}
\def\lb{\boldsymbol{l}}
\def\rb{\boldsymbol{r}}

\def\ub{\boldsymbol{u}}

\def\xb{\boldsymbol{x}}

\def\Fb{\boldsymbol{F}}

\def\Tb{\boldsymbol{T}}

\def\kbT{k_\mathrm{B} T}

\newcommand{\avg}[1]{\left\langle #1 \right\rangle}

\def\<{\langle} \def\>{\rangle}

\DeclareMathOperator{\atantwo}{atan2}

\begin{document}
\title{Microscopic origin of tunable assembly forces in chiral active environments}
\author{Clay H. Batton}
\author{Grant M. Rotskoff}
\email{rotskoff@stanford.edu}
\affiliation{Department of Chemistry, Stanford University, Stanford, CA, USA 94305}
\date{\today}
\date{\today}
\begin{abstract}
    Across a variety of spatial scales, from nanoscale biological systems to micron-scale colloidal systems, equilibrium self-assembly is entirely dictated by---and therefore limited by---the thermodynamic properties of the constituent materials. 
    In contrast, nonequilibrium materials, such as self-propelled active matter, expand the possibilities for driving the assemblies that are inaccessible in equilibrium conditions. 
    Recently, a number of works have suggested that active matter drives or accelerates self-organization, but the emergent interactions that arise between solutes immersed in actively driven environments are complex and poorly understood.
    Here, we analyze and resolve two crucial questions concerning actively driven self-assembly: i) How, mechanistically, do active environments drive self-assembly of passive solutes? ii) Under which conditions is this assembly robust?
We employ the framework of odd hydrodynamics to theoretically explain numerical and experimental observations that chiral active matter, i.e., particles driven with a directional torque, produces robust and long-ranged assembly forces.
Together, these developments constitute an important step towards a comprehensive theoretical framework for controlling self-assembly in nonequilibrium environments. 
\end{abstract}

\maketitle
\section{Introduction} \label{sec:intro}

Synthetic and biological systems that consume energy to produce persistent directed motion have come to be known as ``active matter''.
The nonequilibrium dynamics of active matter can lead to dramatic changes in collective behavior, including motility-induced phase separation in colloidal Janus particles~\cite{cates_motility-induced_2015} and flocking in bacterial swarms~\cite{czirok1996formation}.
The interactions between passive materials immersed in such active matter systems remain poorly understood.  
Many recent works have sought to explore~\cite{frey_self-organisation_2020,ma_inverse_2021,zhang_active_2017,rey_light_2023} and characterize~\cite{li_non-equilibrium_2020,floyd_quantifying_2019,solon_generalized_2018} the consequences of an active environment for self-assembly.
Experimental active matter systems, including Janus particles~\cite{palacci_living_2013} and active dumbbells~\cite{suma_dynamics_2014,cugliandolo_phase_2017}, may provide driving forces that stabilize assemblies of passive particles that, in equilibrium conditions, have no propensity to self-assemble~\cite{stenhammar_activity-induced_2015, mallory_active_2018,dolai2018phase}.
This far-from-equilibrium ``actively driven self-assembly'' offers a compelling route to designing and controlling materials that are not accessible in equilibrium conditions, but the microscopic driving forces are not well understood.

Currently, there is no complete microscopic theory that explains how active matter drives self-assembly; here we aim to close this theoretical gap. 
Our theory is built by understanding the induced interactions between solutes that arise from the fluctuations of an active bath.
The approach we take yields both physical explanations and numerical predictions for effective interactions between solutes in an active bath. 
Our focus is motivated in part by observations from recent experiments~\cite{grober_unconventional_2023} suggesting that active matter with directional, chiral torques can accelerate assembly of passive solutes that otherwise aggregate slowly.

Our theoretical framework requires analyzing the force exerted by an active bath on passive solutes---ultimately the driving force for assembly---which has been studied in a variety of contexts~\cite{solon_pressure_2015, ni2015tunable}.
Perhaps most relevant to our current work, the induced force between two parallel walls that arises from the nonequilibrium fluctuations of an active bath of active Brownian particles (ABPs) has been dubbed an active Casimir effect~\cite{ni2015tunable,ray2014casimir}, though this analogy is misleading.
As we show, the oscillatory force profile for ABPs can be entirely explained by packing effects related to the finite size of the particles (cf. Sec.~\ref{sec:packing}). 
This is not a force that can be explained by density fluctuations in a continuous field at these length scales, as, for example, occurs in classical hydrophobicity~\cite{chandler_gaussian_1993,li_fluctuation-induced_1992}.
To achieve robust assembly, the forces that arise from an active bath must manifest as long-ranged attractive interactions.
We show, theoretically and numerically, that such forces do not arise with ABPs as the solvent. 

In the case of chiral active matter, however, the collective motion of the bath is fundamentally different from ABPs, creating distinct opportunities for modulating interactions between passive particles.
When the particle dynamics breaks chiral symmetry, the corresponding hydrodynamic equations can have diffusivities and viscosities that are antisymmetric tensors, often called ``odd'' hydrodynamics~\cite{banerjee_odd_2017}. 
We show that the emergent odd transport properties that arise from actively driven torques can drive stable and large-scale assembly. 
Microscopically, active particle currents near the passive solutes lead to stable and robust effective assembly forces. 

\begin{figure*}[ht]
    \centering
    \includegraphics[width=1.0\linewidth]{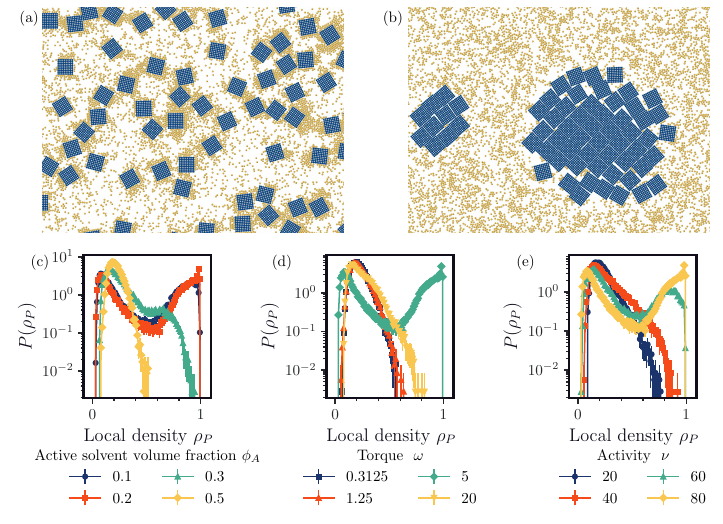}
    \caption{
    Multiple passive squares of length $9\ \sigma$ in a chiral active Brownian particle bath with a passive volume fraction of $ 0.2 $. (a,b) Systems where the passive squares do not and do assemble for $ \phi_{A} = 0.2 $, $ \nu = 80 $, and (a) $ \omega = 0.3125 $ and (b) $ \omega = 5 $, respectively. Histograms of local density of the passive particles, $ \rho_{P} $, for (c) varying $ \phi_{A} $ at $ \omega = 5 $ and $ \nu = 80 $, (d) varying $ \omega $ at $ \phi_{A} = 0.2 $ and $ \nu = 80 $, (e) varying $ \nu $ at $ \omega = 5 $ and $ \phi_{A} = 0.2 $.
    See Sec.~\ref{sec:ld} for further details on how the local density is computed.
    }
    \label{fig:chiral_sa}
\end{figure*}

With straightforward theoretical arguments, a minimal continuum model of odd diffusivity, and extensive numerical simulations, we comprehensively characterize the microscopic origins of assembly forces between passive objects in two and three-dimensional active baths.
We show that for active baths without chiral self-propulsion, attractive forces do not arise except in the true ``Casimir'' regime, the limit of extremely low density. 
Remarkably, chirality, when appropriately tuned, can manifest long-ranged and stable assembly forces for passive particles, as illustrated in Fig.~\ref{fig:chiral_sa}.
What is more, this attraction appears to be driven not by collective fluctuations, but rather by fluxes induced by odd diffusivity.

\paragraph*{Prior work on actively driven self-assembly}

\begin{figure*}[ht]
    \centering
    \includegraphics[width=0.8\linewidth]{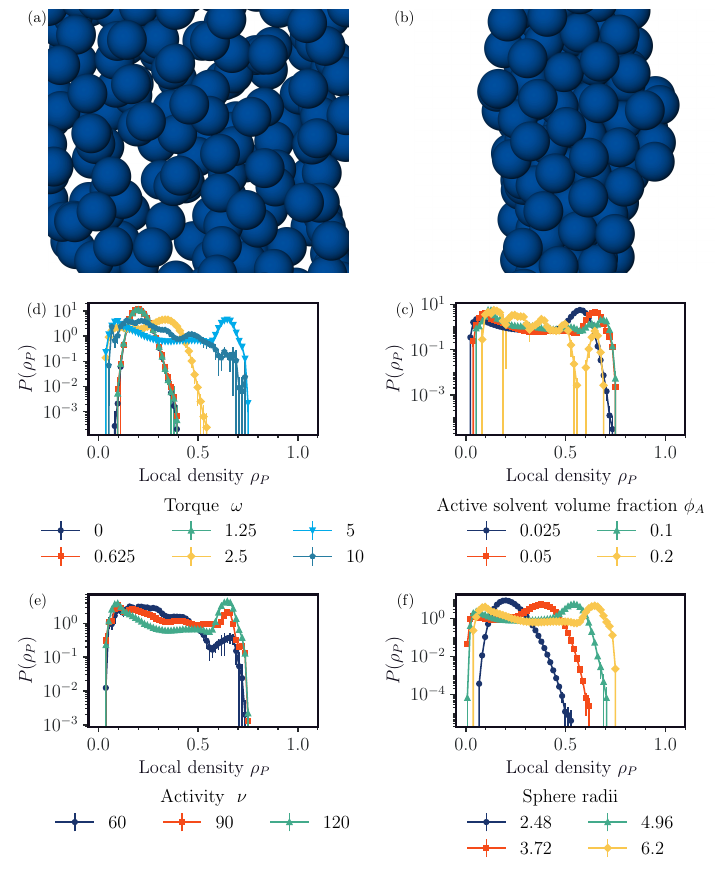}
    \caption{
        Multiple passive spheres of radii $6.2\ \sigma $ in a chiral active Brownian particle bath with a passive volume fraction of $ 0.2 $ in three dimensions. (a,b) Systems where the passive spheres do not and do assemble for $ \phi_{A} = 0.2 $, $ \nu = 120 $, and (a) $ \omega = 0.625 $ and (b) $ \omega = 5 $, respectively, where the solvent is not visualized.
        Histograms of local density of the passive sphere particles in three dimensions, $ \rho_{P} $, for $ \phi_{P} = 0.2 $ at (c) varying $ \omega $ with $ \phi_{A} = 0.05 $, $ \nu = 120 $ and $ r = 6.2\ \sigma $, (d) varying $ \phi_{A} $ at $ \nu = 120 $, $ \omega = 5 $, and $ r = 6.2\ \sigma $, (e) varying $ \nu $ at $ \phi_{A} = 0.05 $, $ \omega = 5 $, and $ r = 6.2\ \sigma $, (f) varying $ r $ at $ \phi_{A} = 0.05 $, $ \nu = 120 $, and $ \omega = 5 $.
    }
    \label{fig:chiral_sa_sphere}
\end{figure*}

There are disparate observations in the literature of active matter driving self-assembly, but no unifying theory has emerged. 
While nonequilibrium self-assembly remains a widely studied topic~\cite{whitelam_statistical_2015}, the works mostly closely related to our investigation here include experiments and simulations by Grober et al.~\cite{grober_unconventional_2023} demonstrate that clustering of sticky passive particles is accelerated by active matter and the emergent structures are strongly modulated by a chiral active Brownian particle bath.
However, their work does not examine the microscopic flows of the bath particles in the vicinity of the passive objects, nor does it consider purely repulsive passive particles, both of which are assessed in our present work.
Similar observations of phase separation induced by chiral active particles were documented numerically in Ref.~\cite{ma_driving_2017}. 
Also closely related, a series of works by Mallory et al.~\cite{mallory_active_2018, mallory_activity-enhanced_2019, mallory_universal_2020} demonstrate that nonequilibrium perturbations arising from an active bath can provide a self-organizing force.
They investigate, for example, a setting in which the active particles are designed with an inherent asymmetry that produces directional flows and hence kinetically induced aggregation~\cite{mallory_active_2018}.
Asymmetries in the passive particles in an active bath have also been found to induce long-range interactions that can lead to assembly~\cite{baek2018generic,granek2020bodies}.
These mechanisms are not as general as those investigated here, where the passive particles do not have any asymmetric interaction with the active bath. 
Yang et al.~\cite{yang2021edge} reported assembly of passive particles driven by a high-density bath of inertial chiral active particles, though the mechanism is not thoroughly characterized.
Assembly of passive particles in a model where the active solvent has a local torque that causes alignment of the active particle orientation with the surface of the passive particles, in comparison to a constant global torque used here, has been observed in simulations, with similar experimental results involving active bacteria and passive colloids~\cite{gokhale2022dynamic,kushwaha2023phase}.
While the model is similar to the one we consider and sees some similar results, namely assembly 
being preferred at large torques and passive particle sizes, it differs in that assembly occurs in an intermediate regime of active velocities and for high solvent densities, along with the mechanism behind the assembly not being fully understood.
While together these results clearly indicate the utility of active matter for accelerating self-assembly dynamics, without a clear microscopic understanding of the driving forces, the ability to design and control assemblies is limited.

\begin{figure}[h!]
    \centering
    \includegraphics[width=0.9\linewidth]{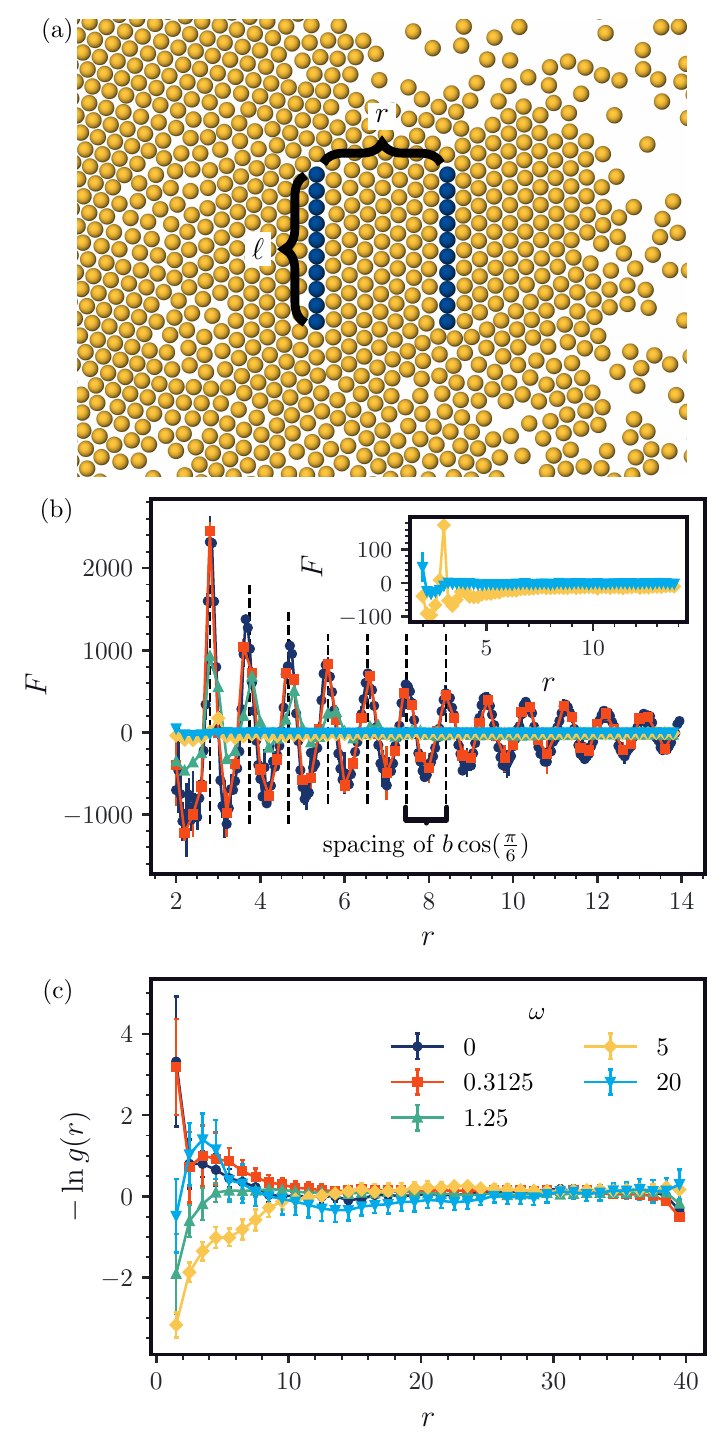}
    \caption{(a) The passive walls in an active Brownian particle bath system previously considered in Refs.~\cite{ray2014casimir,ni2015tunable} for walls of length $ \ell = 10 $, separation $ r = 8 $, $ \rho = 0.4 $, $ \nu = 80 $, and $ \omega = 0 $.  (b,c) The (b) force $ F $ and (c) effective free energy $ - \ln g ( r ) $ on walls of length $ 10 $, $ \rho = 0.4 $, and $ \nu = 80 $ for varying torque $ \omega $. Vertical dashed lines in (b) correspond to the lattice spacing $ b $ of a hexagonal lattice \cite{lauersdorf2021phase}. Error bars in (c) are computed over twelve independent simulations. See Sec.~\ref{sec:ffe} for further details on how the force and free energy are computed.}
    \label{fig:abp_force_profile}
\end{figure}

\section{Robust assembly in a chiral active Brownian particle bath}

In this work we consider solvent and solutes that evolve under overdamped Langevin dynamics for two-dimensional and three-dimensional systems.
The solvent is modeled as chiral active Brownian particles, in which the $i$th solvent particle evolves in two dimensions as \cite{liao2018clustering,ma2022dynamical}
\begin{align} 
    \dot{\rb}_i ( t ) & = \xi^{-1} [\Fb_i + \nu \bb_i ( t )] + \sqrt{2D_t} \Lambdab_{i},                   \\
    \bb_i ( t )    & = [\cos\theta_i ( t ), \sin\theta_i ( t )]^\top, \\
    \dot{\theta}_i ( t ) & = \omega + \sqrt{2D_r} \Gamma_{i} \,,
\end{align}
where $\rb_i$ denotes the position of the $i$th particle, $ \xi $ is the translational drag coefficient, $\Fb_i $ is the force,  $\bb_i$ denotes the direction of its active velocity, $ \nu $ is the magnitude of the active force, $ \omega $ is the active torque, and $ D_t $ and $ D_r $ are the translational and rotational diffusion constants that are related by the formula $ D_r = 3 \ \sigma^{-2} D_t $ and $ D_t $ is related to $ \xi $ by the Stokes-Einstein relation $ D_t = \kbT \xi^{-1} $.
Here, $ \Lambdab_i $ and $ \Gamma_i $ are independent Gaussian white noises with zero mean and unit variance.
The particles interact with the Weeks-Chandler-Anderson (WCA) potential \cite{weeks1971role}, given by the sum $ U = \sum_{i\neq j} u(l_{ij}) $, where $ l_{ij} = \left| \rb_i - \rb_j \right| $ is the distance between particles $ i $ and $ j $.
The form of $ u(l_{ij}) $ is given by
\begin{equation}
    u ( l_{ij} ) = 4 \epsilon_{ij} \left[ \left( \frac{\sigma_{ij}}{l_{ij}} \right)^{12} - \left( \frac{\sigma_{ij}}{l_{ij}} \right)^{6} + \frac{1}{4} \right] \theta \left( 2^{\frac{1}{6}} - \frac{l_{ij}}{\sigma_{ij}} \right) \,, \label{eq:u_wca_2}
\end{equation}
where $ \epsilon_{ij} $ and $ \sigma_{ij} $ are the energy and length scales set by the particle types, respectively, and $ \theta $ is the Heaviside function.
The corresponding force is $ \Fb_i = - \diff{U(t)}{\rb_i} $.
We choose the diffusion constants for the solvent and solutes to be equal with $ D_t = D_{t,b} = 1 $, $ \kbT = 1 $, and $ \epsilon_{ij} = 40 $ and $ \sigma_{ij} = 1 $ unless otherwise specified, along with simulations being performed in two dimensions unless otherwise specified.
Further simulation details, including details for rigid body simulations and three dimensional simulations, are provided in Sec.~\ref{sec:appwca}.

Chiral active matter drives reliable, dynamical assembly of passive solutes, but the microscopic mechanism leading to this phenomenon is utterly distinct from that of the achiral case in which assembly does not occur. 
Fig.~\ref{fig:chiral_sa} illustrates that self-assembly occurs for an appropriate combination of the solvent parameters of torque $\omega$, activity $\nu$, and active solvent volume fraction $\phi_A$.
We quantify the assembly by constructing histograms of the local density of passive solutes $\rho_P$; phase separation corresponds to the coexistence of a low density and high density region (see Sec.~\ref{sec:ld} for further details). 
As shown in Figs.~\ref{fig:chiral_sa} (c-e), a range of parameters support self-assembly when the torque is sufficiently large (see also Fig.~\ref{fig:voronoi_phase_diagram}), along with larger square size driving assembly as shown in Fig.~\ref{fig:voronoi_analysis_size}.
Furthermore, assembly does not depend strongly on the shape or dimensionality of the object. 
In Fig.~\ref{fig:chiral_sa_sphere}, we quantify the propensity of passive spherical objects to assemble in 3D, which occurs over a range of active torques, activities, solvent densities, and passive sphere radii.
Similar trends hold for passive cubes immersed in a bath of chiral active particles in 3D (Fig.~\ref{fig:chiral_sa_cube}).
We similarly studied passive disks (Fig.~\ref{fig:chiral_sa_circ}) and triangular passive particles (Fig.~\ref{fig:chiral_sa_tri}) to ensure that observations in 2D were not narrowly tailored to square geometries, where the analysis in terms of the inter-wall forces is most physically transparent.

\subsection{Force and free energy profiles in a model system}

To assess the underlying molecular fluctuations governing this activity-induced self-assembly, we consider the minimal model depicted in Fig.~\ref{fig:abp_force_profile} (a).
In this geometry, we consider two parallel walls of length $l$ separated by a distance $r$ in a bath of chiral active particles.
We define the total interaction force as
\begin{equation}
    F_{\rm wall}^{({\rm tot})}[\rho(\xb)] = F_{\rm wall}^{({\rm int})}[\rho(\xb)] - F_{\rm wall}^{({\rm ext})}[\rho(\xb)]
\end{equation}
using the sign convention that if the force applied to the walls by particles in the interstitial region exceeds the force applied by particles outside this region, then the force is positive, and the walls will repel.
The force is computed over a range of torques in Fig.~\ref{fig:abp_force_profile} (b).
For achiral ABPs, this force oscillates between attractive and repulsive, as previously reported by Ni et al.~\cite{ni2015tunable}.
For sufficiently large torques, the oscillatory force profile is not maintained and a long-ranged attractive force sets in, though with a much smaller magnitude. 

To further assess the propensity for passive solutes to self-assemble in given nonequilibrium bath conditions, we compute the effective nonequilibrium free energy profile for the solute degrees of freedom. 
To do so, we draw inspiration from liquid state theory~\cite{chandler1983review}, and quantify the effective interaction by measuring the radial distribution function, or $g(r)$ for fluctuating passive solutes. 
Our solutes are non-spherical and hence evolve dynamically as rigid bodies composed of smaller particles (see Sec.~\ref{sec:appwca} for further details).
The radial distribution function is computed between the centers of mass of the passive solutes.
In Fig.~\ref{fig:abp_force_profile} (c), we show the effective interaction for chiral active particles over a range of torques.
We plot $-\ln g(r)$ because, in equilibrium, this quantity would correspond to the reversible work required (or gained) when bringing two solutes into contact. 
Away from equilibrium, this thermodynamic interpretation is no longer valid, but a positive value of $-\ln g(r)$ as $r\to 0$ indicates that it is statistically unlikely for the two solutes to come together.
For ABPs, this is the case, with repulsion at close distances being observed and holding across densities except for weak attraction at low density (Fig.~\ref{fig:free_energy_density}), leading to assembly not being dynamically accessible in most conditions.
In contrast, for the chiral case, there is a long-ranged attractive interaction over a range of torques that drives assembly.

\subsection{Local density profile depends strongly on torque}

\begin{figure}
    \centering
    \includegraphics[width=0.9\linewidth]{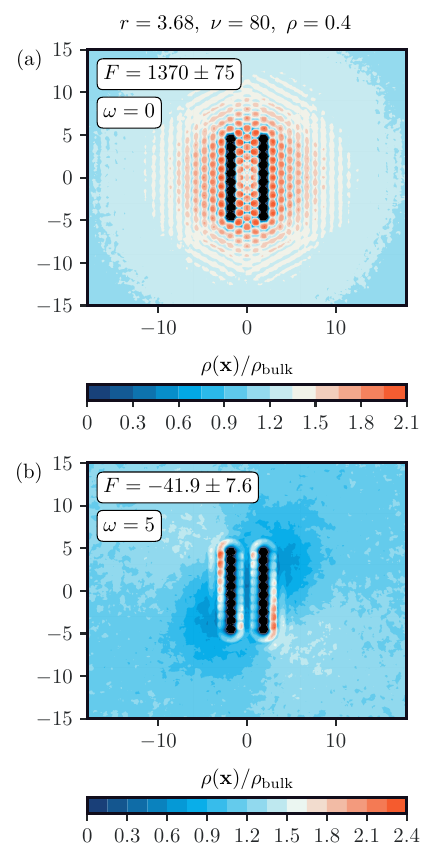}
    \caption{Density and orientation fields for maximal repulsive force between two passive walls in a chiral active Brownian particle bath at (a) $ \omega = 0 $ and (b) $ \omega = 5 $. The orientation field is \emph{not} scaled by the magnitude of the average local current, but only indicates its direction.}
    \label{fig:gdo_0_summary}
\end{figure}

The nature of the force profiles can be understood by examining the local density and orientation fields of the chiral active particles.
In the achiral case, the oscillatory force profile between two walls results from changes in the typical local density at differing separation distances.
Microscopically, the enhancement in local density both between and outside the walls results from the slow orientational relaxation of achiral active particles. 
In Fig.~\ref{fig:gdo_0_summary} (a), we computed the density and orientation fields for $\omega=0$ at a separation of $3.68\ \sigma$, a value at which the force is maximal. 
The average orientations of the particles in the region of enhanced density point towards the boundary of the passive object.
This trend is robust across separation distances and force magnitudes, as shown in Figs.~\ref{fig:gd_0}-\ref{fig:go_4} (a). 
Mechanistically, the force generation depends strongly on large local density enhancements. 
The nature of the density enhancements rely on entropic packing effects, and from this a minimal model of the force profile in the achiral case can be constructed (see Sec.~\ref{sec:packing} for further details).

\begin{figure*}[ht]
    \includegraphics[width=0.8\linewidth]{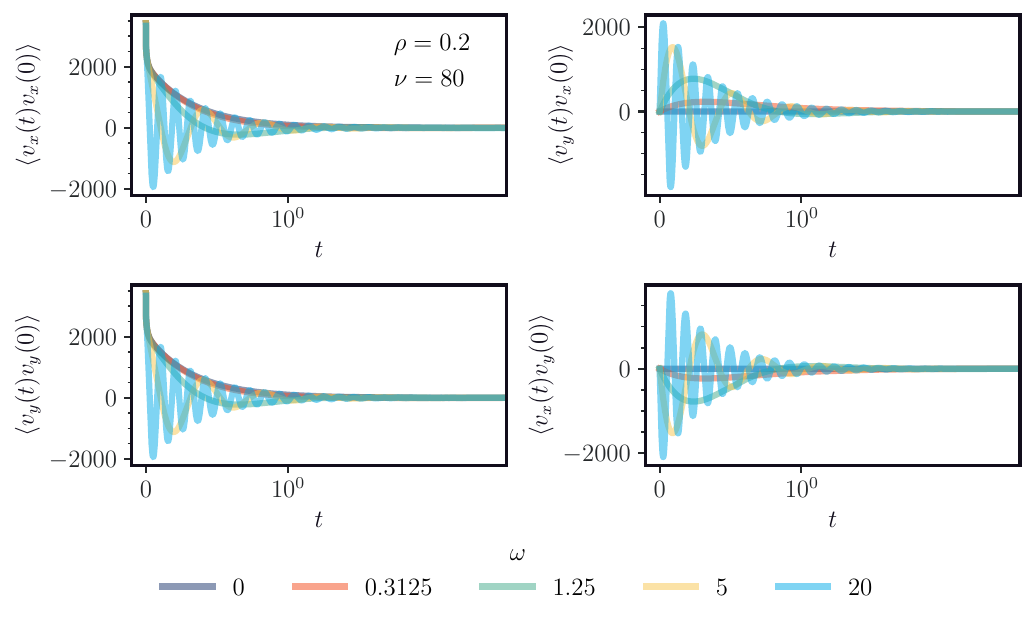}
    \caption{Autocorrelation functions computed to determine $\mathsf{D}_{\rm a}$ from Eq.~\eqref{eq:gkint}.}
    \label{fig:greenkubo}
\end{figure*}

In comparison, chiral motion does not lead to a long-lived density enhancement at the boundary of a passive wall, and, in fact, local microscopic fluctuations are much more subtle. 
Due to the nonzero torque, chiral active particles do not simply aggregate at the boundary of a passive object, but rather flow parallel to the boundary. 
This boundary flux is evident in Fig.~\ref{fig:gdo_0_summary} (b), where $\omega=5$, but also for all values of $\omega$ that we tested (cf. Figs.~\ref{fig:gd_0}-\ref{fig:go_4}).
For sufficiently small torques relative to the active velocity, the chiral active particles produce small boundary fluxes and have a force profile that is similar in magnitude and shape to that of the ABPs, as shown in~\ref{fig:abp_force_profile} (b).
The similarity in the density and orientation fields for small torques ($\omega=0.3125$) is evident in Figs.~\ref{fig:gd_0}-\ref{fig:gd_4} (b) and Figs.~\ref{fig:go_0}-\ref{fig:go_4} (b).

For large torques, the fluxes at the boundary of a passive object are sufficiently large that density does not accumulate proximal to the passive object (Figs.~\ref{fig:gd_0}-\ref{fig:gd_4} (e)). 
As a result, force generation is consistently near zero for $\omega=20$, as shown in Fig.~\ref{fig:abp_force_profile} (b).
In this regime, no assembly occurs, as shown in Fig.~\ref{fig:chiral_sa} and quantified by the effective free energy profile in Fig.~\ref{fig:abp_force_profile} (c).

In the intermediate regime, the interplay between boundary fluxes and density accumulation can lead to robust assembly forces. 
Fig.~\ref{fig:chiral_sa} demonstrates that assembly does indeed occur when $\omega=5$.
This particular set of conditions for the chiral active Brownian particle bath leads to a long-ranged attractive interaction, which decays over roughly 15 particle diameters (Fig.~\ref{fig:abp_force_profile} (c)). 
Microscopically, it is evident from numerical simulations that the density accumulates asymmetrically, with more particles on the outside compared to the region between the two walls (Figs.~\ref{fig:gdo_0_summary} (b), ~\ref{fig:wall_density_torque_trends}, and ~\ref{fig:wall_rho_hex_torque_80}).
This phenomenon results in higher applied forces on the outside, driving the passive objects together. 

\section{Odd diffusivity drives assembly}

\begin{figure*}[ht]
    \includegraphics[width=0.85\linewidth]{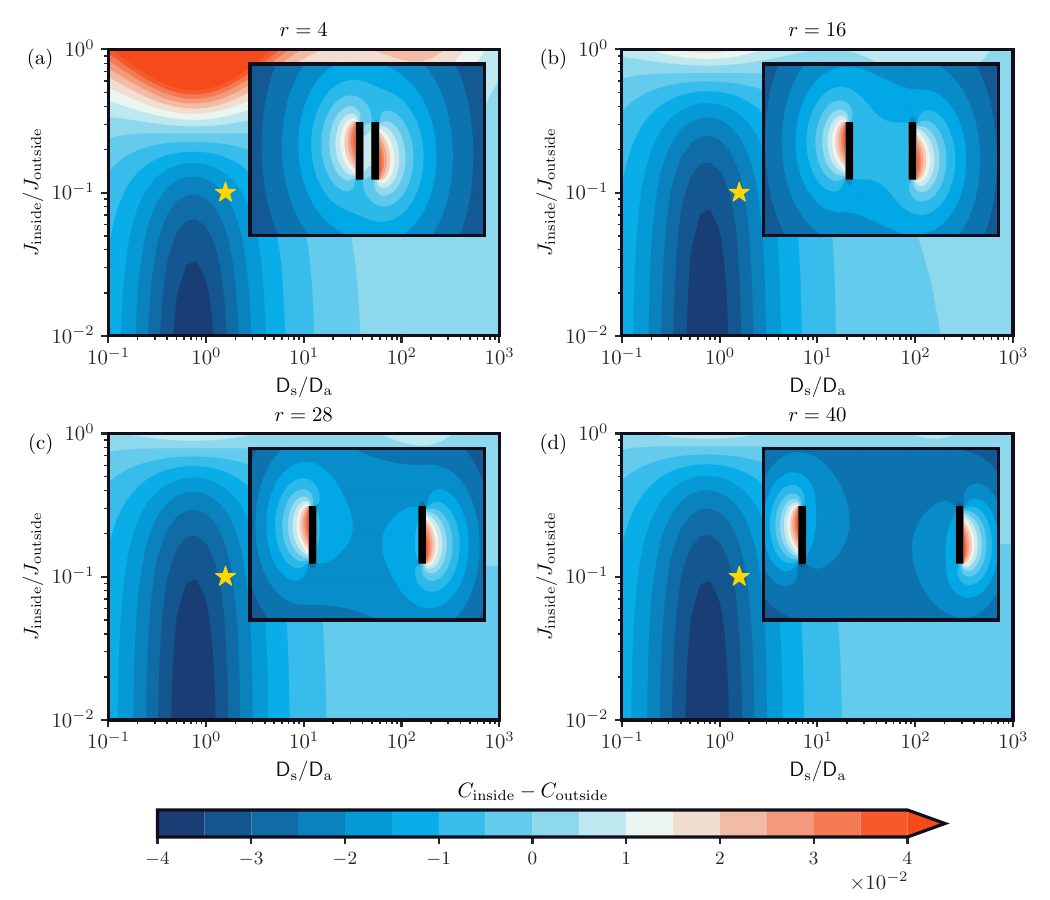}
    \caption{Difference between the concentrations near the walls internally and externally for varying ratios of $ \mathsf{D}_{\rm s} / \mathsf{D}_{\rm a} $ and the internal and external fluxes at the walls for different separation distances, $ r $, in each panel, evaluated using a finite difference scheme where the fluxes parallel and perpendicular to the walls are discretized at second and fourth order, respectively. The external flux is set to $ J_{\mathrm{outside}} = 10 \mathsf{D}_{\rm a} \mathsf{D}_{\rm s}^{-1} $ to maintain a relatively similar range of concentrations across the parameters. Inset images correspond to the concentration profiles for the parameters at the gold star.}
    \label{fig:odd_diffusion}
\end{figure*}

To assess the microscopic origins of the asymmetric density field and hence the attractive assembly forces for this narrow range of torque values, we construct a minimal model of the concentration profile. 
Chiral active liquids break time-reversal symmetry, due to implicit energy consumption by the particles, and parity with their single particle torques and lead to ``odd'' hydrodynamic response functions~\cite{banerjee_odd_2017}.
These nonequilibrium liquids have anomalous transport properties, including odd diffusion tensors and stress tensors~\cite{liao_mechanism_2019}.
Most relevant to our setting, Hargus et al.~\cite{hargus_odd_2021} showed that chiral active particles can be modeled by a simple continuum description of ``odd diffusivity''. 
The implications of an odd diffusion tensor manifest only in the presence of a boundary that breaks translational symmetry for the active bath, such as the presence of passive particles. 
In continuum models, this has been shown to both enhance diffusivity and create directional particle fluxes~\cite{kalz_collisions_2022a}. 

\begin{figure}
    \centering
    \includegraphics[width=0.8\linewidth]{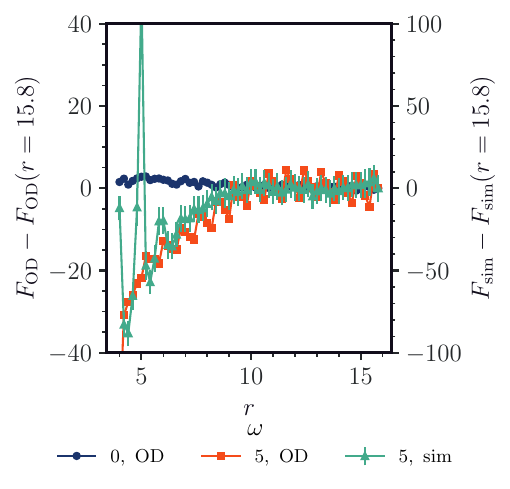}
    \caption{Force measured from finite difference scheme $ F_{\mathrm{OD}} $, in which the fluxes parallel and perpendicular to the walls are expanded to second and fourth order respectively for walls of length $ 10\ \sigma $, width $ 3\ \sigma $, and $ \omega = 0 $ and $ \omega = 5 $, and the force measured from simulations, $ F_{\mathrm{sim}} $, for $ \omega = 5 $.
    Force offsets correspond to $ F_{\mathrm{OD}} (r ; \omega = 0) = 1.0$, $ F_{\mathrm{OD}} ( r ; \omega = 5) = 16.1 $, and $ F_{\mathrm{sim}} ( r ; \omega = 5) = -12 \pm 8 $, with the simulation results corresponding to $ \rho = 0.2 $ and $ \nu = 80 .$}
    \label{fig:odd_diff_forces}
\end{figure}

Each passive particle leads to a non-vanishing steady-state mass current in its proximity.
Fig.~\ref{fig:gdo_0_summary} shows both density and the average orientation for achiral ABPs (a) and chiral ABPs (b). 
In the chiral case, that is, for all $\omega > 0$, there is a net flux parallel to the walls, oriented in opposite $y$-directions on the $-x$ and $+x$ sides. 
Because there are no sources or sinks in our periodic system, the steady state density profile must be consistent with a divergence-free current due to the conservation of mass.

In two dimensions, the active particle density $\rho(x,y)$ satisfies a continuity equation
\begin{equation}
    \partial_t \rho(x,y) = \nabla \cdot \left( \mathsf{D} \cdot \nabla \rho(x,y)\right)
    \label{eq:odd_diff}
\end{equation}
where $\mathsf{D}_{ij} = \mathsf{D}_{\rm s}\delta_{ij} - \mathsf{D}_{\rm a} \varepsilon_{ij}$, $\delta_{ij}$ is a Kronecker $\delta$-function, and $\varepsilon_{ij}$ is the antisymmetric Levi-Civita tensor. 
Without boundary conditions, the continuity equation results in a steady-state concentration profile that is independent of $\mathsf{D}_{\rm a}$; however, when the density has Neumann boundary conditions, the antisymmetric contribution to the diffusion tensor can affect the steady-state. 
We note that similar results hold in three dimensions, but require specifying the direction of the active torque in order to determine the form of the diffusion tensor. 

To determine if odd diffusivity plays a significant role in shaping the density field of the chiral active particles, we numerically solve~\eqref{eq:odd_diff} in conditions where assembly does and does not occur. 
To do so, we first estimated the flux around the boundary of a passive object by computing
\begin{equation}
    \Jb (x,y) = \frac{1}{\xi T} \sum_{i=1}^{N} \int_0^T (\Fb_i + \nu \bb_i ( t )) k\bigl((x,y), \rb_i\bigr) dt,
\end{equation}
where the integral is over a simulation of duration $T$ in the steady state.
We use a Gaussian kernel (described in detail in Sec.~\ref{sec:field_details}) to obtain a smooth flux field. 

The diffusion coefficients can be directly related to the velocities via a Green-Kubo relation~\cite{hargus_odd_2021}, 
\begin{equation}
    \begin{aligned}
        \mathsf{D}_{\rm s} &= \frac{1}{2} \int_0^\infty \avg{ v_i(t) v_j(0) } \delta_{ij} dt, \\
        \mathsf{D}_{\rm a} &= -\frac{1}{2} \int_0^\infty \avg{ v_i(t) v_j(0) } \varepsilon_{ij} dt, \\
    \end{aligned}
    \label{eq:gkint}
\end{equation}
where $ \avg{\dots} $ denotes an average over all solvent particles, and $ v_i(t) $ is the $ i $-th component of the velocity of a particular solvent particle at time $ t $.
We compute these integrals with numerical quadrature after obtaining the time correlation functions using the Wiener-Khinchin theorem.
These autocorrelation functions are shown in Fig.~\ref{fig:greenkubo}, with diffusion coefficients across $ \rho $ and $ \omega $ shown in Fig.~\ref{fig:diffusion_trend}.
The oscillatory autocorrelation functions that we observe are consistent with recent theoretical work for odd diffusive dynamics~\cite{kalz_oscillatory_2023}.
With the boundary fluxes and the diffusion tensor determined, we then solve for the steady-state density profile using finite differences.
We describe the numerical details in Sec.~\ref{sec:numerical_details}.

As shown in Fig.~\ref{fig:odd_diffusion}, we compute the stationary density profile as a function of the ratio of the flux in the interstitial region $J_{\rm inside}$ to the flux on the outside boundary $J_{\rm outside}$ and also the ratio of the symmetric to antisymmetric part of the diffusion tensor, $\mathsf{D}_{\rm s}/\mathsf{D}_{\rm a}$.
We then computed the difference between the average concentration near the walls in the interstitial and a region of the same area on the outside; this is a proxy for the magnitude of the induced attractive force because the total force exerted on the walls is proportional to the local concentration. 
At short separation distances, we see (Fig.~\ref{fig:odd_diffusion} (a-b)) that this concentration difference is negative over a large range of $J_{\rm inside}/J_{\rm outside}$, provided that $\mathsf{D}_{\rm a}$ is appreciable relative to $\mathsf{D}_{\rm s}$.
As the separation between the walls grows, this effect persists but becomes considerably weaker, as correlations between the adjacent walls in the interstitial region decay, as shown in Fig.~\ref{fig:odd_diffusion} (c-d).
The inset concentration profiles in Fig.~\ref{fig:odd_diffusion} are in good qualitative agreement with the density profiles obtained from direct numerical simulation in Figs.~\ref{fig:gd_0}-\ref{fig:gd_4}.

\subsection{Minimal model of odd diffusivity captures effective attraction}

While the difference in concentration that we plot in Fig.~\ref{fig:odd_diffusion} is highly suggestive of an attractive force, to quantify the resulting force, we develop a simple mean-field model of the force profile using the computed concentration profiles.
The excess force into the wall can be computed, for a steady state density profile $\rho_{\rm ss}(\xb)$, as
\begin{equation}
    F_{\rm wall}[\rho_{\rm ss}(\xb)] = \int_{\textrm{wall}} [F(\xb) + v \bb(\xb)] \cdot \hat{\eb}_{1}\ \rho_{\rm ss}(\xb) d\xb,
\end{equation}
which is simply the projection of the total force onto the wall and $\xb\equiv (\rb, \theta)$; the unit vector $\hat{\eb}_1$ is aligned perpendicular to the wall.
This force accounts for particle-particle interactions in addition to the wall-particle interactions.   
Making a mean-field assumption that the excess force into the wall arises not from inter-particle interactions, but instead from the larger scale particle flows and the active velocity, we obtain
\begin{equation}
    F_{\rm wall}^{(\textrm{int})} \approx \int_{\textrm{int}} \bar f(\rb) \cdot \hat{\eb}_1\  \rho_{\rm ss}(\rb) d\rb
\end{equation}
with the active velocity $ \bar f(\rb) $ being found from the total active velocity $ \nu $ and the active velocity parallel to the wall $ \sigma J_{y} $ per
\begin{equation}
    \bar f(\rb) = \sqrt{\nu^2 - \sigma^2 J_y^2(\rb)}.
\end{equation}
This expression allows us to estimate the force on the walls using the steady-state density obtained by solving~\eqref{eq:odd_diff} with Neumann flux boundary conditions, imposed using the numerically measured fluxes. 
Integrating this expression over the regions ``int'' and ``ext'', defined to be a rectangle of area $\ell \times \sigma$ immediately abutting the wall on the interior and exterior, respectively,  we obtain numerical values for $F_{\textrm{wall}}^{(\textrm{int})}$ and $F_{\textrm {wall}}^{(\textrm{ext})}$.
As shown in Fig.~\ref{fig:odd_diff_forces}, this minimal model captures the correct magnitude of the force and also is in good agreement with its spatial range.
The agreement is due to the two-body nature of the interactions between the solvent and the passive objects at higher torque values as the density of active particles at passive object boundaries is lowered, as shown in Figs.~\ref{fig:chiral_sa}(a,b) and~\ref{fig:gdo_0_summary}.
This assumption breaks down at low separation distances, at which point the confined space between the walls leads to more complex interactions leading to the spikes in the simulation force profile as shown in Fig.~\ref{fig:odd_diff_forces}.

\section{Conclusion}

Both experimental~\cite{grober_unconventional_2023} and computational~\cite{ni2015tunable, mallory_activity-enhanced_2019} studies have emphasized the potentialities of an active environment for accelerating and controlling self-assembly.
However, the microscopic dynamics of the active matter environment have a profound influence and determine whether or not assembly is possible.
Our work here demonstrates that a striking difference emerges when active baths break chiral symmetry: while achiral active particles do not drive assembly, chiral active matter robustly and stably drives aggregation of passive solutes. 
Remarkably, this phenomenon occurs even when those passive solutes have no attractive interaction amongst themselves.
In other words, the propensity to self-assemble results entirely from the fluctuations of the chiral active solvent.

Further extensions of this work could focus on several themes, namely further development of the theory, and the ability to attain certain structures through incorporation of more complex interactions and protocols.
While odd diffusion provides an adequate description of the solvent interaction with the solutes, extensions of prior work looking at interactions of ideal active Brownian particles with passive objects~\cite{wagner2022steady} could be used to provide a more detailed understanding of the forces that drive assembly.
In particular, active field theories could be used to describe from a first-principles approach the interactions between the chiral active solvent and the passive solutes in a manner that captures both interacting solvent particles and the effects of chirality~\cite{bruna2022phase,kalz_fieldtheory_2023}.
Additionally, this work has considered only WCA interactions, with the incorporation of other potentials leading to potentially interesting assembly outcomes~\cite{rechtsman2005optimized,rechtsman2006designed,rechtsman2006self}.
In conjunction with recent work considering how to design nonequilibrium protocols to reach desired structures~\cite{chennakesavalu2021probing,chennakesavalu2024adaptive}, the use of chiral active matter to drive self-assembly could be a powerful tool for the design of materials.

We extensively characterize the differences between chiral and achiral active solvents both numerically and theoretically. 
We provide clear evidence that the robust assembly we observe occurs in both two and three dimensions, depends weakly on the shape and size of the solute, and remains robust over a range of densities and activities. 
We provide a complete theoretical explanation for the origins of self-assembly forces in these far-from-equilibrium environments.
The forces that arise between solutes in active Brownian particle systems can be explained entirely by density effects, as shown in Sec.~\ref{sec:packing}, and we show that these forces do not drive assembly in any regime.
Chiral active solvents, however, lead to odd transport dynamics and flows of the active solvent near the solutes lead to long range and attractive forces. 
Taken together, these results provide a proscriptive theory for understanding self-assembly in complex, active environments. 
Our results lay the foundations to characterize and ultimately control actively driven self-assembly in both biological and synthetic environments.

\section*{Conflicts of interest}

There are no conflicts to declare.

\section*{Acknowledgements}
The authors thank Huiting Liu for helpful discussions. 
This material is based upon work supported by the U.S. Department of Energy, Office of Science, Office of Basic Energy Sciences, under Award Number DE-SC0022917. This research used resources of the National Energy Research Scientific Computing Center (NERSC), a U.S. Department of Energy Office of Science User Facility located at Lawrence Berkeley National Laboratory, operated under Contract No. DE-AC02-05CH11231, and the Sherlock cluster, operated by Stanford University and the Stanford Research Computing Center.

\paragraph*{Data and Code Availability:}
Code and input scripts for all simulations are available on GitHub \texttt{\href{{https://github.com/rotskoff-group/tunable-assembly}}{https://github.com/rotskoff-group/tunable-assembly}}.

\bibliographystyle{apsrev4-2}
\bibliography{bib}

\onecolumngrid
\appendix
\clearpage
\newpage

\setcounter{figure}{0}
\makeatletter 
\renewcommand{\thefigure}{S\@arabic\c@figure}
\makeatother

\section{Analyzing the force between solutes in an achiral active Brownian particle bath}
\label{sec:packing}

\begin{figure*}[h]
    \centering
    \includegraphics[width=1.0\linewidth]{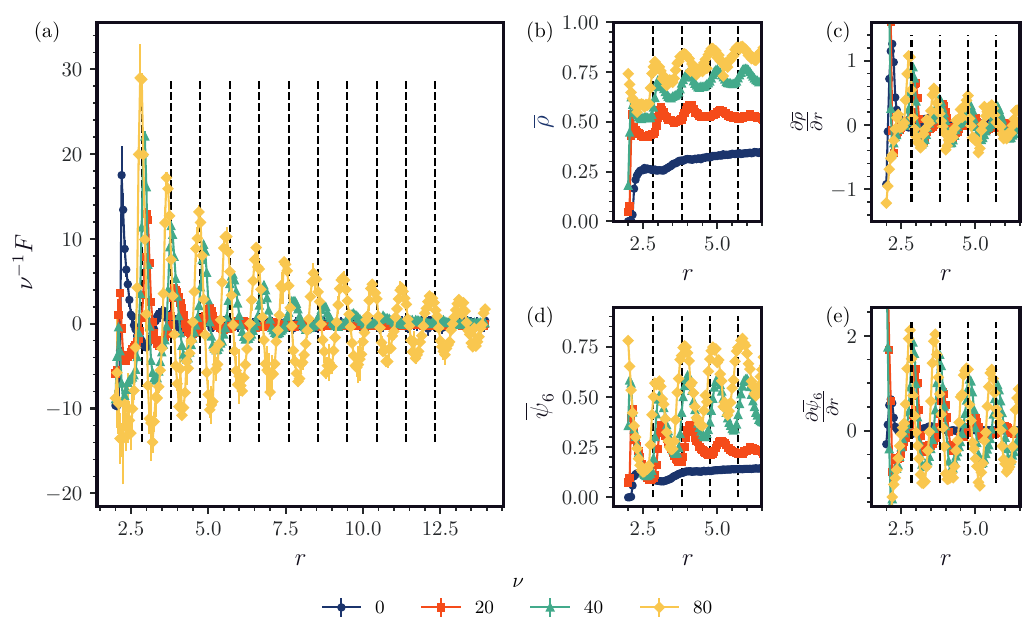}
    \caption{Quantities obtained for a system with two passive walls of length $ 10 $ and $ \rho = 0.4 $ for varying $ \nu $ and separation length. (a) Force $ F $. (b) Density between walls, $ \overline{\rho} $. (c) Hexatic order parameter between walls, $ \frac{\partial \overline{\rho}}{\partial r} $. (d) $ \overline{\psi}_6 $. (e) $ \frac{\partial \overline{\psi}_6}{\partial r} $. See Sec.~\ref{sec:dh} for further details on how the density and hexatic order parameter are computed.}
    \label{fig:wall_force_rho_hex}
\end{figure*}

\begin{figure}[h]
    \centering
    \includegraphics[width=1.0\linewidth]{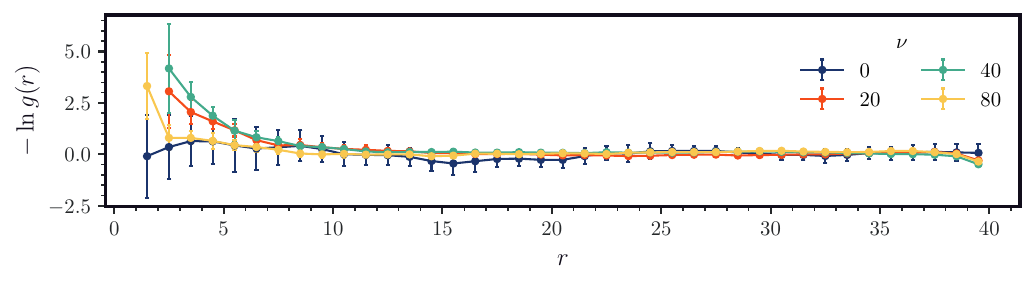}
    \caption{The effective free energy $ - \ln g ( r ) $ between two passive walls of length $ 10 $ and $ \rho = 0.4 $ for varying activity.}
    \label{fig:free_energy_non_chiral}
\end{figure}

\begin{figure*}[h]
    \centering
    \includegraphics[width=1.0\linewidth]{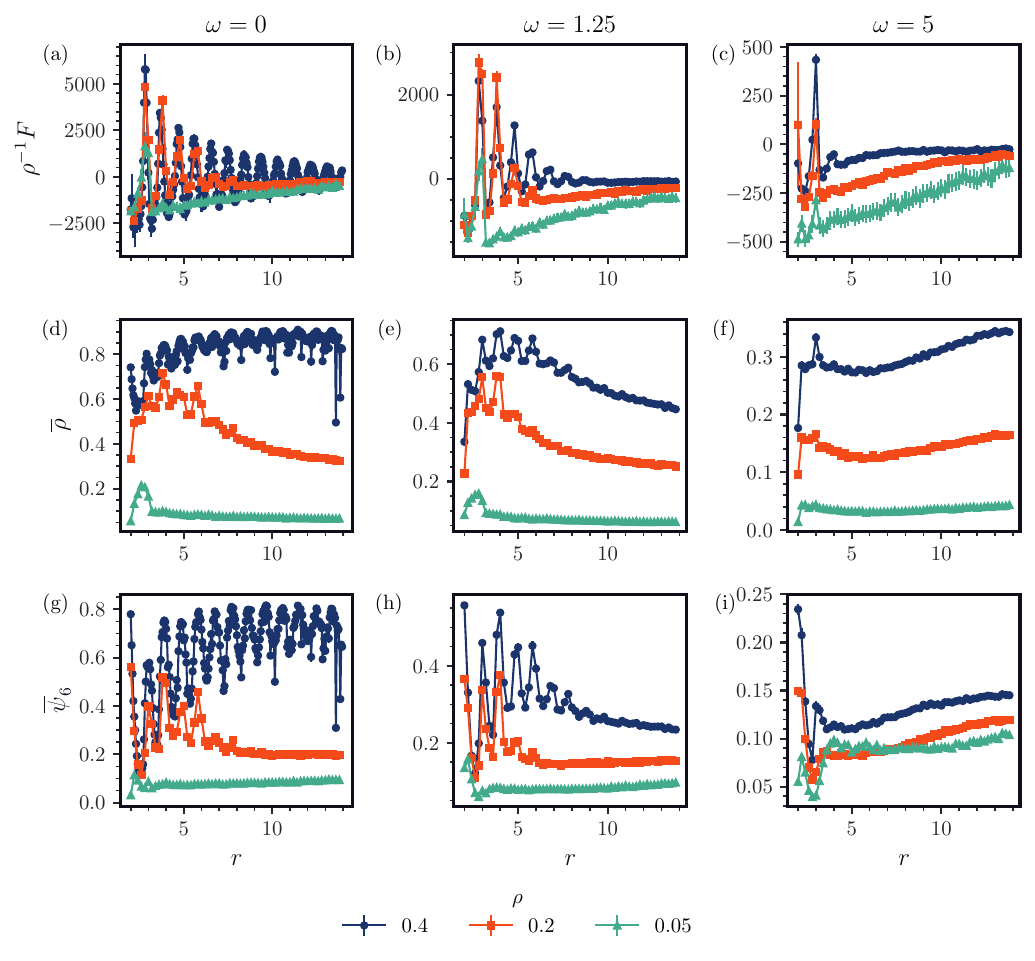}
    \caption{$ \rho^{-1} F $, $ \overline{\rho} $, and $ \overline{\psi}_6 $ between two passive walls of length $ 10 $ for varying separation lengths, torque, and $ \rho $ for $ \nu = 80 $. See Sec.~\ref{sec:dh} for further details on how the density and hexatic order parameter are computed.}
    \label{fig:wall_density_torque_trends}
\end{figure*}

The dominant contribution to forces both internal and external for a passive solute in an achiral active Brownian particle bath is the local enhancement of density around the solute (see Fig.~\ref{fig:abp_force_profile} (b)).
The large density gradient perpendicular to the walls forms through a mechanism similar to motility-induced phase separation (MIPS): particles orient into the direction of the wall and generate a force in proportion to the average local density a distance $\sigma$ (given by the WCA potential) away from the wall, which we denote $\rho_{\rm wall}$.
This force can be approximated as $F(\rho_{\rm wall}) \approx \nu l \sigma \tau \rho_0$ where $\tau$ is the characteristic rotational diffusion time.
When the walls are separated by a distance larger than the length over which the density is enhanced, $F_{\rm wall}^{({\rm int})} = -F_{\rm wall}^{({\rm ext})}$ and the interaction vanishes.
In fact, any attractive force for achiral (and nearly achiral) active particles arises from density correlations in the interstitial region.
This basic picture holds over a variety of conditions, including different points in the MIPS phase diagram (i.e., different choices of total density and active velocity) as shown in Figs.~\ref{fig:wall_force_rho_hex}-\ref{fig:wall_density_torque_trends}.

A minimal model illustrates that the oscillatory force profile for achiral particles arises entirely due to packing constraints: the separation distances $r$ at which the two walls accommodate a high-density hexatic packing lead to large repulsive forces, while the separation distances $r$ that are not commensurate with a hexatic lattice lead to smaller values of $F_{\rm wall}^{(\rm int)}$, and consequently attractive forces.
We verified that hexatic order was correlated with the location of the repulsive peaks in the force profile (Figs.~\ref{fig:wall_force_rho_hex} and~\ref{fig:wall_density_torque_trends}) by computing an average of the hexatic order parameter
\begin{equation}
    \psi_6(\rb_k) = \frac16 \sum_{l\in \mathcal{N}(\rb_k)} e^{i 6 \theta_{kl}}, \label{eq:hexatic}
\end{equation}
where $\mathcal{N}(\rb)$ denotes the set of the six nearest neighbors of the particle at position $\rb$ and $\theta_{kl}$ is the angle between the vector $\rb_l-\rb_k$ and $\boldsymbol{e}_x$.
We denote by $\bar\psi_6$ the average value of $\psi_6$ restricted to the region between the two parallel walls.

\begin{figure}[h]
    \centering
    \includegraphics{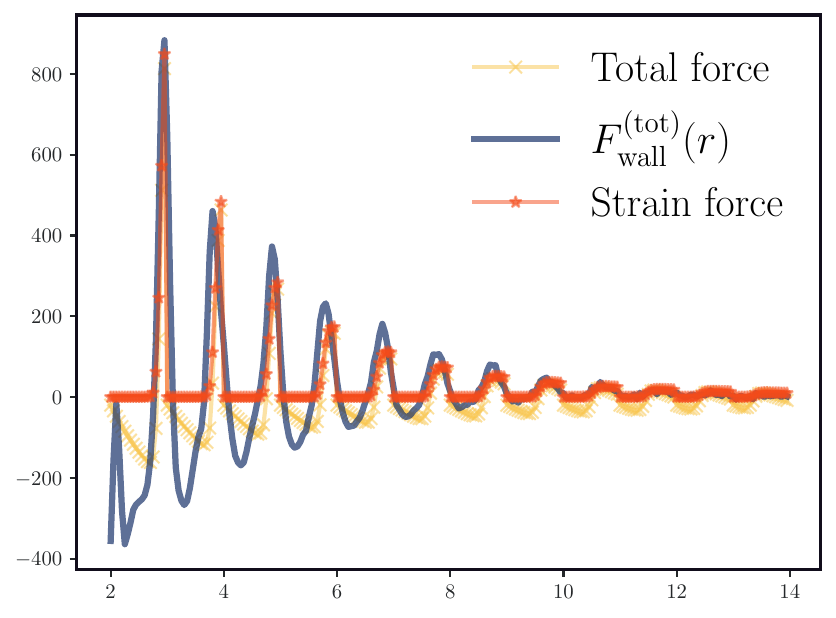}
    \caption{The strain force \eqref{eq:strain} with $\alpha=20$ and $c=0.4$. These parameters were chosen for agreement with the magnitude of the force, but the location of the peaks in the strain force depends weakly on these free parameters.
    }
    \label{fig:abp_force_model}
\end{figure}

The total force on the walls arises from active particles oriented into the walls aggregating and pushing inward.
The forces generated by the active particles balance exactly when the separation between the walls is large: it is inter-particle correlations in the interstitial region that lead to the nontrivial force profile shown in Fig.~\ref{fig:abp_force_profile} (b).
To capture the nature of the forces arising from these inter-particle correlations, we first define $\bar \rho(r)$ as the average density in the interstitial region as a function of the wall separation distance.
At the P\'eclet numbers, given by $\nu \sigma / D_t$, we consider here, the density adjacent to the internal and external walls remains close to the hexatic density.
A minimal model for the density in the interstitial region is then
\begin{equation}
    \bar{\rho}(r) = \frac{n_{\rm hex}(r) + \Delta n_{\rm WCA}(r)}{r \ell }
\end{equation}
where $n_{\rm hex}(r)$ is simply the number of particles accommodated by a hexatic packing $\lfloor 2r/(\sqrt{3} \sigma) \rfloor \ell$ and
\begin{equation}
    \Delta n_{\rm WCA}(r) = \frac{\ell e^{-r D_{r}/\nu}}{1+\exp \left[ - \alpha ( \delta r - \delta_0 r)  \right]}
\end{equation}
with $\delta r = r - \lfloor 2r/(\sqrt{3} \sigma) \rfloor \ell $, an offset $\delta_0 = c (2^{1/6}-1)\sigma$, and a parameter $\alpha$ that accounts for the softness of WCA interaction.
The exponential decay accounts for the decay in correlations with the boundary, and the rate is chosen to be the persistence length for ABPs.
The principal contribution to the average internal force $F^{(\rm int)}_{\rm wall}$ at first order in $r$ is a repulsive inter-particle force due to a strained hexatic packing.
That is, by compressing the space available to the hexatic lattice, the particles in the interstitial region are strained. 
This additional contribution can be calculated easily, 
\begin{equation}
\Delta F^{(\rm int)}_{\rm wall}(r) \approx 18 \sqrt[3]{4} \epsilon \sigma l (\bar \rho(r) - \rho_{\rm hex})^2 
\label{eq:strain}
\end{equation}
which uses a second-order expansion of the repulsive interaction.
This model, despite its strong simplifying assumptions, predicts the location and magnitude of the oscillatory repulsive peaks with surprising accuracy (Fig.~\ref{fig:abp_force_model}).
For separation distances that are just above those that accommodate a hexatic lattice, depletion of achiral particles in the interstitial region leads to lower force generation in $F_{\rm wall}^{(\rm int)}$, and the walls are pushed together by the external active Brownian particle bath. 
This complex interplay of packing and finite-size effects \emph{does not} yield a robust assembly force.

\begin{figure*}[h]
    \centering
    \includegraphics[width=1.0\linewidth]{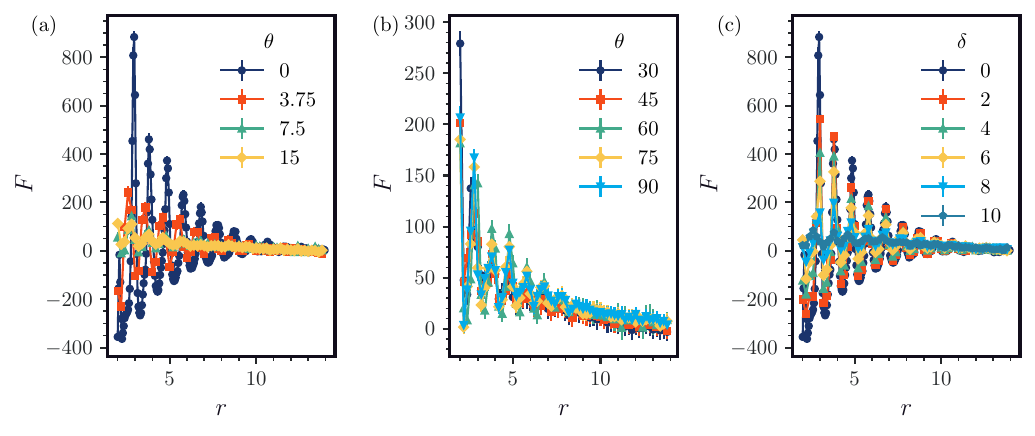}
    \caption{Force between two passive walls of length $ 10 $, $ \rho = 0.4 $, and $ \nu = 40 $ for (a,b) varying angles $ \theta $ between the walls and (c) varying offset $ \delta $ between the walls.}
    \label{fig:wall_force_angle_offset}
\end{figure*}

\begin{figure*}
    \centering
    \includegraphics[width=1.0\linewidth]{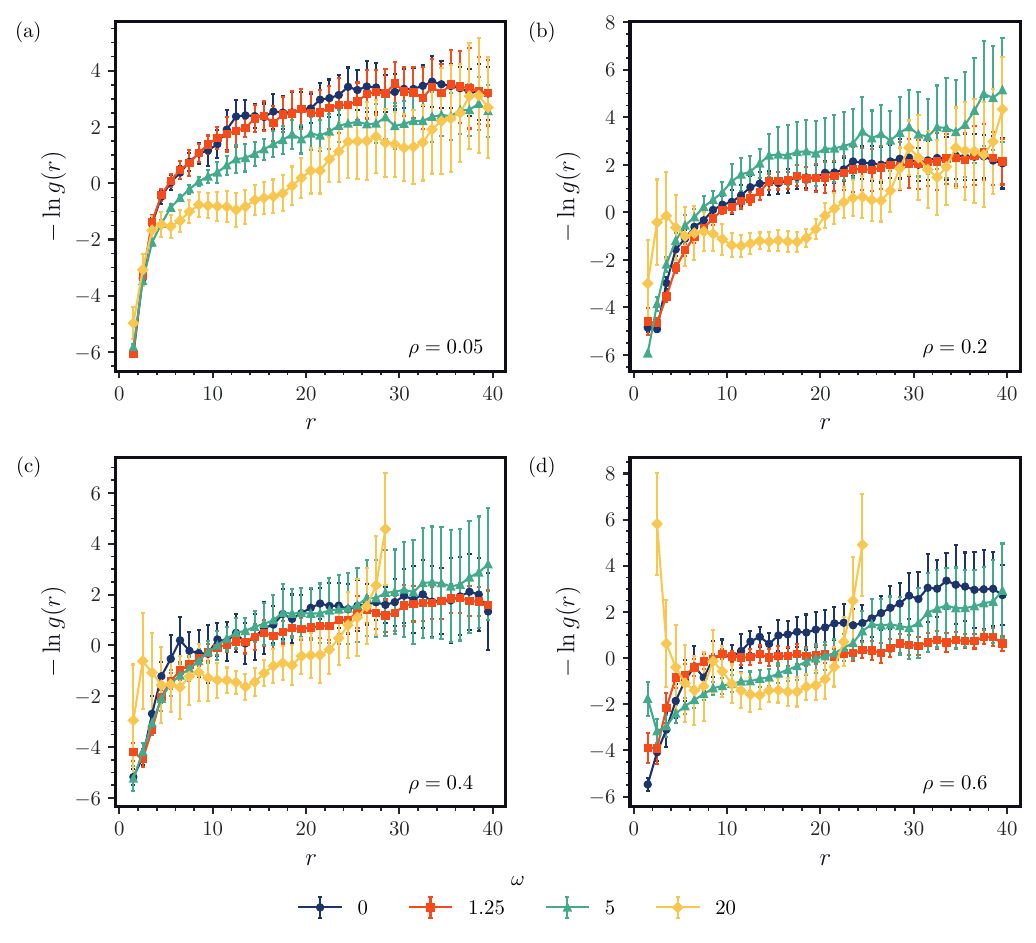}
    \caption{Effective free energy between two passive walls of length $ 10 $ and $ \nu = 80 $ for varying $ \rho $ and $ \omega $ that are constrained to move only in the $ x $-direction and to not rotate. See Sec.~\ref{sec:cs} for further details on how the constraint is implemented.}
    \label{fig:free_energy_constrain}
\end{figure*}

To test the hypothesis that the repulsive force emerges from hexatic order, we conducted extensive numerical simulations under conditions in which this order could not emerge. 
First, at very low densities, there is not a sufficiently large relative enhancement of the local density to achieve a fully packed interstitial region, leading to a $\bar\psi_6 \approx 0$ for all wall separation distances. 
As shown in~Fig.~\ref{fig:wall_density_torque_trends}, there is no oscillatory force and, in fact, there is only a weakly attractive long-ranged force. 
Additionally, we varied the angle between the plates and offset the distance between the center of the plates in the $ y $-direction, processes which disrupt the ordering between the plates, and found in both cases the force between the plates to be diminished (Fig.~\ref{fig:wall_force_angle_offset}).
In contrast, constraining the motion of the plates to be only in the $ x $-direction and hence preserving the hexatic lattice is found to lead to attractive effective free energies between the plates (Fig.~\ref{fig:free_energy_constrain}).

\begin{figure*}
    \centering
    \includegraphics[width=1.0\linewidth]{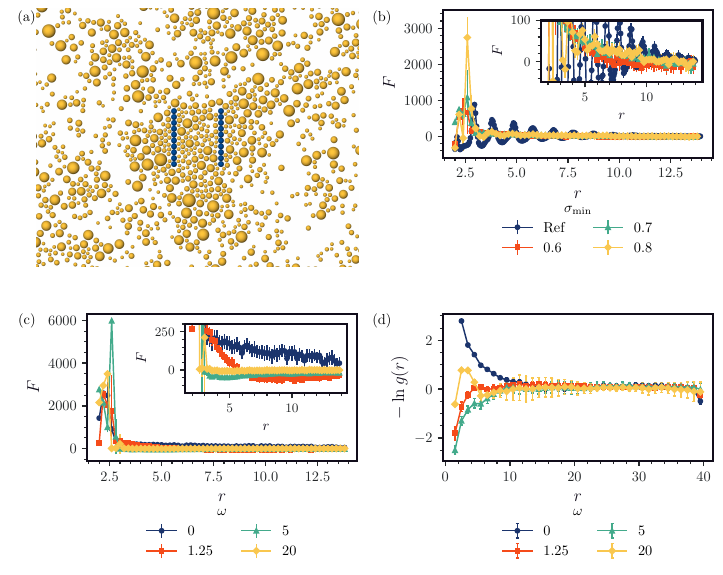}
    \caption{Passive walls in a polydisperse active Brownian particle bath system. (a) Image of the system with $ \rho = 0.4 $, $ \nu = 40 $, and $ \sigma_{\mathrm{min}} = 0.6 $. (b) Force between two passive walls of length $ 10 $, $ \rho = 0.4 $, $ \nu = 40 $, and $ \omega = 0 $ for varying separation lengths and $ \sigma_{\mathrm{min}} $. (c) Force and (d) $ -\ln g ( r ) $ between two passive walls of length $ 10 $, $ \rho = 0.4 $, $ \nu = 80 $, and $ \sigma_{\mathrm{min}} = 0.6 $ for varying separation lengths and $ \omega $.}
    \label{fig:wall_force_glass}
\end{figure*}

At higher particle densities, we disrupted hexatic order by examining a system of continuously polydisperse active particles.
The particle diameters were drawn with a power law decay $P(\sigma) = A \sigma^{-3}$.
This model has been studied in the literature on glassy dynamics due to its resistance to crystallization, even when deeply quenched~\cite{ninarello2017models}. 
Without regular order, the system does not admit the high-density interstitial packings achieved when the bath contains only particles of a single diameter.
Consistent with the minimal model, this effect eliminates the oscillations in the force profile and the effective nonequilibrium free energy $-\ln g(r)$ shows only a short-ranged repulsion as $r\to 0$ (Fig.~\ref{fig:wall_force_glass}; see Sec.~\ref{sec:glass} for further details).

\section{Odd diffusion results}

\subsection{Odd Diffusion Scheme Details} \label{sec:numerical_details}

We solve the steady state form of~\eqref{eq:odd_diff} using a finite difference scheme for a system with two walls.
Far from the walls, the density is assumed to be the bulk density $ \rho_{0} $.
Around the walls, Neumann boundary conditions are imposed by the fluxes, which take the form
\begin{align}
    J_{x} &= - \mathsf{D}_{\rm{s}} \frac{\partial \rho}{\partial x} + \mathsf{D}_{\rm{a}} \frac{\partial \rho}{\partial y} \,, \\
    J_{y} &= - \mathsf{D}_{\rm{a}} \frac{\partial \rho}{\partial x} - \mathsf{D}_{\rm{s}} \frac{\partial \rho}{\partial y} \,,
\end{align}
where the flux perpendicular to the wall is zero by no-flux boundary conditions, and the flux parallel to the wall will be an input based on simulation data.
Under steady-state conditions, ~\eqref{eq:odd_diff} becomes
\begin{equation}
    \mathsf{D}_{\rm{s}} \nabla^{2} \rho = 0 \,, \label{eq:odd_diff_ss}
\end{equation}
which we solve by discretizing $ \rho $ and the differential operators on a grid with equal spacing $ h $ in the $ x $ and $ y $ directions, yielding $ \rho_{i,j} $ in the bulk by a central difference scheme
\begin{equation}
    \mathsf{D}_{\rm{s}} \frac{\rho_{i+1,j} + \rho_{i-1,j} + \rho_{i,j+1} + \rho_{i,j-1} - 4 \rho_{i,j}}{h^{2}} = 0 \,.
\end{equation}
The Neumann boundary conditions are enforced by introducing fictitious grid points at the surface of the walls.
We discretize the derivatives in either $ J_{x} $ and $ J_{y} $ to fourth and second order in the directions perpendicular and parallel to the wall, which yield the expressions
\begin{align}
    -\mathsf{D}_{\rm{s}} \frac{-\rho_{i+2,j}+8\rho_{i+1,j} - 8\rho_{i-1,j} + \rho_{i-2,j}}{12 h} + \mathsf{D}_{\rm{a}} \frac{\rho_{i,j+1} - \rho_{i,j-1}}{2h} &= 0 \,, \\ 
    -\mathsf{D}_{\rm{a}} \frac{-\rho_{i+2,j}+8\rho_{i+1,j} - 8\rho_{i-1,j} + \rho_{i-2,j}}{12 h} - \mathsf{D}_{\rm{s}} \frac{\rho_{i,j+1} - \rho_{i,j-1}}{2h} &= J_{y} \,,
\end{align}
for the areas to the left and right of the walls, with analogous expressions for the areas above and below the walls.
It is found that expanding the perpendicular direction derivative to fourth order for both $ J_{x} $ and $ J_{y} $ leads to a singular matrix, and hence we expand to fourth order for only one of the fluxes.
The fictitious nodes are eliminated by solving for them in terms of the equations given by fluxes and inputting the subsequent values into the governing equation~\eqref{eq:odd_diff_ss}, in which the derivative in the direction to the wall is also expanded to fourth order, to yield a linear system of equations that can be solved for $ \rho_{i,j} $.
In the case of nodes on the left side of the walls in which we expand the fluxes parallel to the walls, $ J_{y} $, to fourth order and the fluxes perpendicular to the walls, $ J_{x} $, to second order, the fictitious nodes are given by
\begin{align}
    -\mathsf{D}_{\rm{s}} \frac{\rho_{i+1,j} - \rho_{i-1,j}}{2 h} + \mathsf{D}_{\rm{a}} \frac{\rho_{i,j+1} - \rho_{i,j-1}}{2h} &= 0 \,, \\ 
    -\mathsf{D}_{\rm{a}} \frac{-\rho_{i+2,j}+8\rho_{i+1,j} - 8\rho_{i-1,j} + \rho_{i-2,j}}{12 h} - \mathsf{D}_{\rm{s}} \frac{\rho_{i,j+1} - \rho_{i,j-1}}{2h} &= J_{y} \,,
\end{align}
which yields for the fictitious nodes $ \rho_{i+1,j} $ and $ \rho_{i+2,j} $
\begin{align}
    \rho_{i+1,j} &= \rho_{i-1,j} + \frac{\mathsf{D}_{\rm{a}}}{\mathsf{D}_{\rm{s}}} \left( \rho_{i,j+1} - \rho_{i,j-1} \right) \,, \\
    \rho_{i+2,j} &= \frac{12 h}{\mathsf{D}_{\rm{a}}} J_{y} + \rho_{i-2,j} + \left(\frac{6 \mathsf{D}_{\rm{s}}}{\mathsf{D}_{\rm{a}}} + \frac{8 \mathsf{D}_{\rm{a}}}{\mathsf{D}_{\rm{s}}} \right) \left( \rho_{i,j+1} - \rho_{i,j-1}  \right) \,,
\end{align}
and upon input to the governing equation, we obtain
\begin{equation}
    \begin{aligned}
        \mathsf{D}_{\rm{s}} h^{-2} &\left[ \left( \rho_{i,j-1} - 2 \rho_{i,j} + \rho_{i,j+1} \right) + \left( -\frac{1}{12} \rho_{i-2,j} + \frac{4}{3} \rho_{i-1,j} - \frac{5}{2} \rho_{i,j} + \frac{4}{3} \rho_{i+1,j} - \frac{1}{12} \rho_{i+2,j} \right) \right] \\
                                   &= -\frac{J_{y}}{h \mathsf{D}_{\rm{a}}} - \frac{1}{6 h^{2}} \rho_{i-2,j} + \frac{8}{3 h^{2}} \rho_{i-1,j} + \frac{6+\frac{3 \mathsf{D}_{\rm{s}}}{\mathsf{D}_{\rm{a}}}-\frac{4 \mathsf{D}_{\rm{a}}}{\mathsf{D}_{\rm{s}}}}{6 h^{2}} \rho_{i,j-1} - \frac{9}{2 h^{2}} \rho_{i,j} + \frac{6-\frac{3 \mathsf{D}_{\rm{s}}}{\mathsf{D}_{\rm{a}}}+\frac{4 \mathsf{D}_{\rm{a}}}{\mathsf{D}_{\rm{s}}}}{6 h^{2}} \rho_{i,j+1} = 0 \,.
    \end{aligned} \label{eq:left_par4perp2_2}
\end{equation}
A similar procedure is followed for the nodes on the right, top, and bottom sides of the walls.
Rewriting~\eqref{eq:left_par4perp2_2} slightly, we obtain for the governing equations around the walls
\begin{align}
    &\begin{aligned} \label{eq:left_par4perp2}
        \textrm{Left:} \ \mathsf{D}_{\rm{s}} &\mathsf{D}_{\rm{a}} \rho_{i-2,j}-16 \mathsf{D}_{\rm{s}} \mathsf{D}_{\rm{a}} \rho_{i-1,j}+\left(-6 \mathsf{D}_{\rm{s}} \mathsf{D}_{\rm{a}}-3 \mathsf{D}_{\rm{s}}^2+4 \mathsf{D}_{\rm{a}}^2\right) \rho_{i,j-1}+27 \mathsf{D}_{\rm{s}} \mathsf{D}_{\rm{a}} \rho_{i,j}\\
                      &+\left(-6 \mathsf{D}_{\rm{s}} \mathsf{D}_{\rm{a}}+3 \mathsf{D}_{\rm{s}}^2-4 \mathsf{D}_{\rm{a}}^2\right) \rho_{i,j+1}+6 h \mathsf{D}_{\rm{s}} J_y = 0 \,,
	\end{aligned} \\ 
    &\begin{aligned} \label{eq:right_par4perp2}
        \textrm{Right:} \ &-\left(6 \mathsf{D}_{\rm{s}} \mathsf{D}_{\rm{a}}-3 \mathsf{D}_{\rm{s}}^2+4 \mathsf{D}_{\rm{a}}^2\right) \rho_{i,j-1} + 27 \mathsf{D}_{\rm{s}} \mathsf{D}_{\rm{a}} \rho_{i,j}-16 \mathsf{D}_{\rm{s}} \mathsf{D}_{\rm{a}} \rho_{i+1,j}+\mathsf{D}_{\rm{s}} \mathsf{D}_{\rm{a}} \rho_{i+2,j} \\
                                                &\quad-\left(6 \mathsf{D}_{\rm{s}} \mathsf{D}_{\rm{a}}+3 \mathsf{D}_{\rm{s}}^2-4 \mathsf{D}_{\rm{a}}^2\right) \rho_{i,j+1}-6 h \mathsf{D}_{\rm{s}} J_y=0 \,, 
	\end{aligned} \\ 
    &\begin{aligned} \label{eq:bottom_par4perp2}
        \textrm{Bottom:} &-\left(6 \mathsf{D}_{\rm{s}} \mathsf{D}_{\rm{a}}-3 \mathsf{D}_{\rm{s}}^2+4 \mathsf{D}_{\rm{a}}^2\right) \rho_{i-1,j}+\mathsf{D}_{\rm{s}} \mathsf{D}_{\rm{a}} \rho_{i,j-2}-16 \mathsf{D}_{\rm{s}} \mathsf{D}_{\rm{a}} \rho_{i,j-1}+27 \mathsf{D}_{\rm{s}} \mathsf{D}_{\rm{a}} \rho_{i,j}\\
        &\quad-\left(6 \mathsf{D}_{\rm{s}} \mathsf{D}_{\rm{a}}+3 \mathsf{D}_{\rm{s}}^2-4 \mathsf{D}_{\rm{a}}^2\right) \rho_{i+1,j}-6 h \mathsf{D}_{\rm{s}} J_x=0 \,,
	\end{aligned} \\ 
    &\begin{aligned} \label{eq:top_par4perp2}
        \textrm{Top:} &\left(-6 \mathsf{D}_{\rm{s}} \mathsf{D}_{\rm{a}}-3 \mathsf{D}_{\rm{s}}^2+4 \mathsf{D}_{\rm{a}}^2\right) \rho_{i-1,j} + 27 \mathsf{D}_{\rm{s}} \mathsf{D}_{\rm{a}} \rho_{i,j} +\left(-6 \mathsf{D}_{\rm{s}} \mathsf{D}_{\rm{a}}+3 \mathsf{D}_{\rm{s}}^2-4 \mathsf{D}_{\rm{a}}^2\right) \rho_{i+1,j}\\
                      &\quad-16 \mathsf{D}_{\rm{s}} \mathsf{D}_{\rm{a}} \rho_{i,j+1}+\mathsf{D}_{\rm{s}} \mathsf{D}_{\rm{a}} \rho_{i,j+2}+6 h \mathsf{D}_{\rm{s}} J_x=0 \,.
	\end{aligned}  
\end{align}

\begin{figure*}[ht]
    \includegraphics[width=0.85\linewidth]{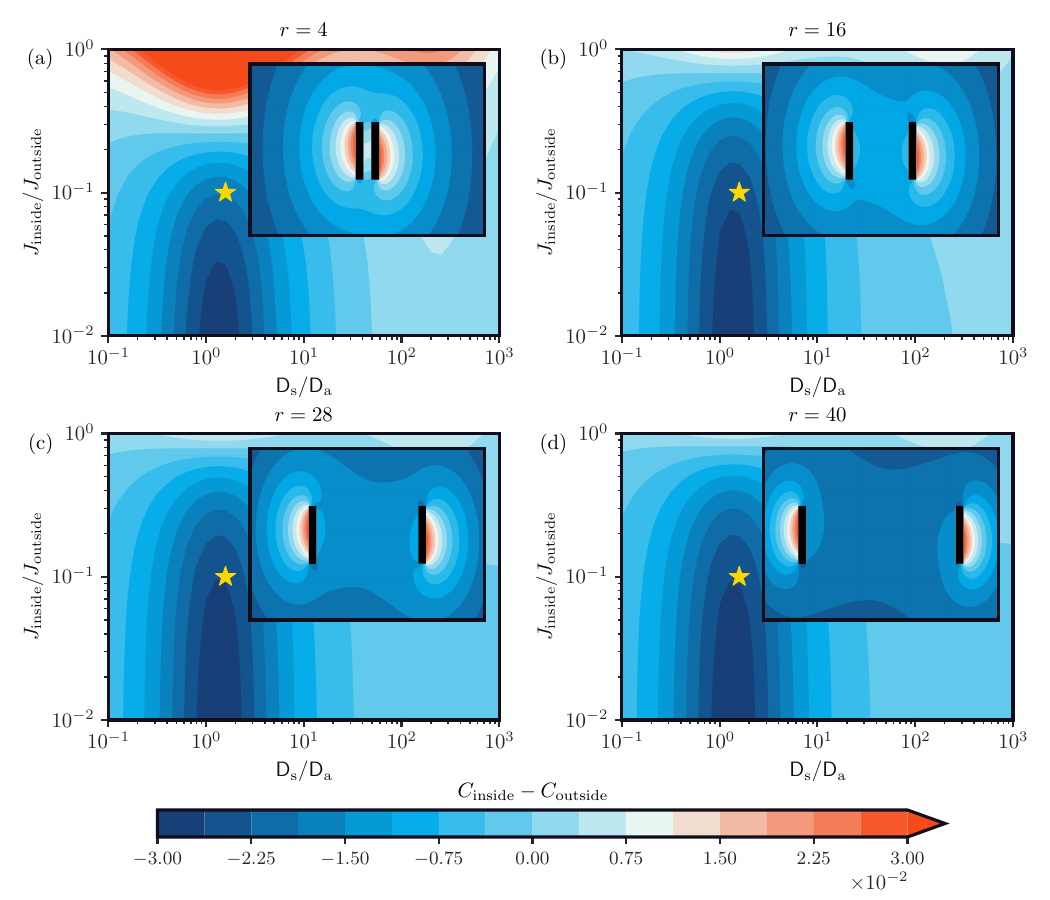}
    \caption{Difference between the concentrations near the walls internally and externally for a varying ratio of $ \mathsf{D}_{\rm s} / \mathsf{D}_{\rm a} $ and the internal and external fluxes at the walls for different separation distances, $ r $, in each panel, evaluated using a finite difference scheme where the fluxes parallel and perpendicular to the walls are discretized at fourth and second order, respectively. The external flux is set to $ J_{\mathrm{outside}} = 10 \mathsf{D}_{\rm a} \mathsf{D}_{\rm s}^{-1} $ to maintain a relatively similar range of concentrations across the parameters. Inset images correspond to the concentration profiles for the parameters at the gold star.}
    \label{fig:odd_diffusion_2}
\end{figure*}

\begin{figure}
    \centering
    \includegraphics[width=0.4\linewidth]{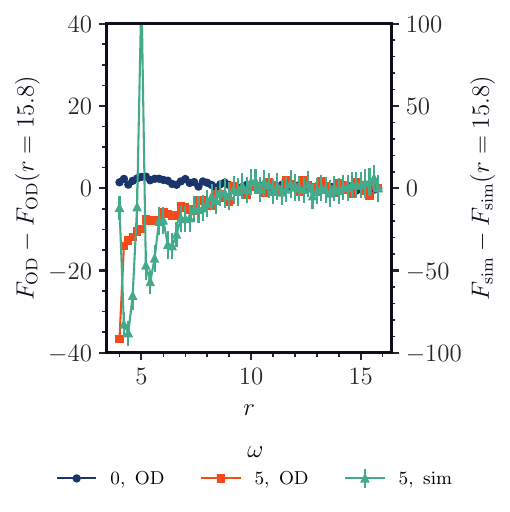}
    \caption{Force measured from finite difference scheme $ F_{\mathrm{OD}} $, in which the fluxes parallel and perpendicular to the walls are expanded to fourth and second order respectively for walls of length $ 10\ \sigma $, width $ 3\ \sigma $, and $ \omega = 0 $ and $ \omega = 5 $, and the force measured from simulations, $ F_{\mathrm{sim}} $, for $ \omega = 5 $.
    Force offsets correspond to $ F_{\mathrm{OD}} (r ; \omega = 0) = 1.1$, $ F_{\mathrm{OD}} ( r ; \omega = 5) = 7.1 $, and $ F_{\mathrm{sim}} ( r ; \omega = 5) = -12 \pm 8 $, with the simulation results corresponding to $ \nu = 80 $ and $ \rho = 0.2 $.}
    \label{fig:odd_diff_forces_2}
\end{figure}

\begin{figure}
    \centering
    \includegraphics[width=0.8\linewidth]{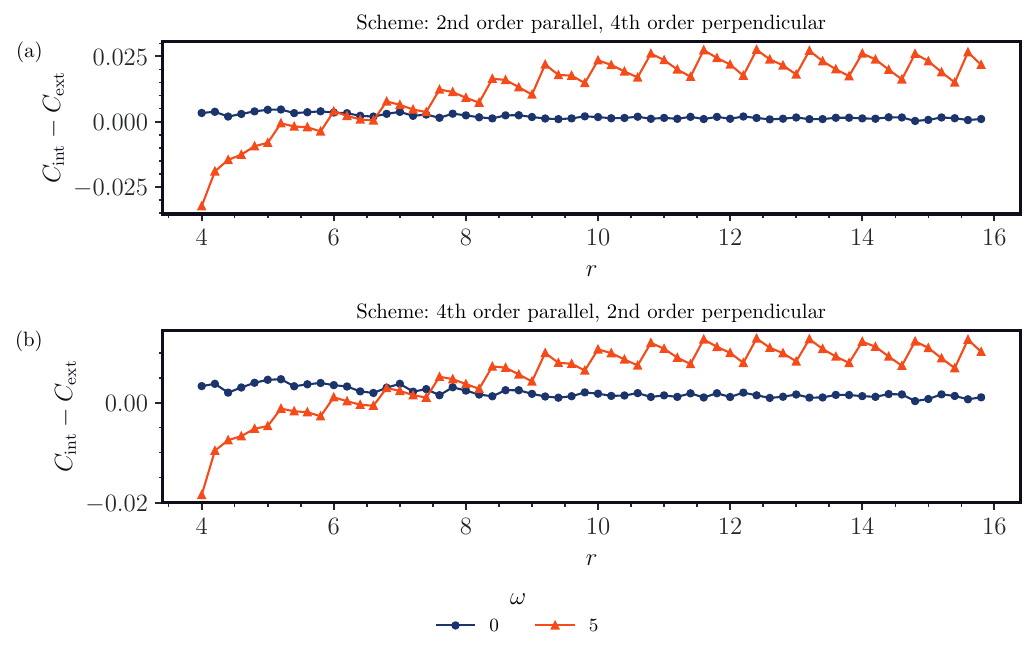}
    \caption{Concentration difference between the internal and external regions of the walls for varying schemes with parameters obtained from simulations at varying $ \omega $ and fixed $ \nu = 80 $ and $ \rho = 0.2 $.}
    \label{fig:odd_diff_conc}
\end{figure}

\begin{figure*}
    \centering
    \includegraphics[width=1.0\linewidth]{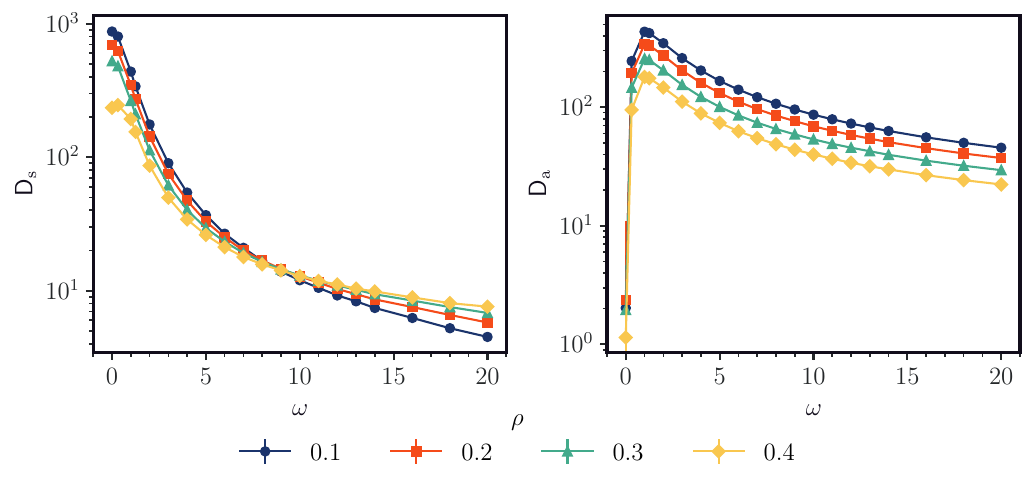}
    \caption{Diffusion constants for varying $ \omega $ and $ \rho $ at $ \nu = 80 $.}
    \label{fig:diffusion_trend}
\end{figure*}

Alternatively, one can expand the fluxes parallel to the walls to second order and the fluxes perpendicular to the walls to fourth order.
In this case, the fictitious nodes on the left side of the walls are given by
\begin{align}
    -\mathsf{D}_{\rm{s}} \frac{-\rho_{i+2,j}+8\rho_{i+1,j} - 8\rho_{i-1,j} + \rho_{i-2,j}}{12 h} + \mathsf{D}_{\rm{a}} \frac{\rho_{i,j+1} - \rho_{i,j-1}}{2h} &= 0 \,, \\ 
    -\mathsf{D}_{\rm{a}} \frac{\rho_{i+1,j} - \rho_{i-1,j}}{2 h} - \mathsf{D}_{\rm{s}} \frac{\rho_{i,j+1} - \rho_{i,j-1}}{2h} &= J_{y} \,,
\end{align}
which yields for the fictitious nodes $ \rho_{i+1,j} $ and $ \rho_{i+2,j} $
\begin{align}
    \rho_{i+1,j} &= -\frac{2 h}{\mathsf{D}_{\rm{a}}} J_y + \rho_{i-1,j} + \frac{\mathsf{D}_{\rm{s}}}{\mathsf{D}_{\rm{a}}} \left( \rho_{i,j+1} - \rho_{i,j-1} \right) \,, \\
    \rho_{i+2,j} &= -\frac{16 h}{\mathsf{D}_{\rm{a}}} J_{y} + \rho_{i-2,j} - \left(\frac{8 \mathsf{D}_{\rm{s}}}{\mathsf{D}_{\rm{a}}} + \frac{6 \mathsf{D}_{\rm{a}}}{\mathsf{D}_{\rm{s}}} \right) \left( \rho_{i,j+1} - \rho_{i,j-1}  \right) \,,
\end{align}
and we obtain for the governing equation
\begin{equation}
    \begin{aligned}
        \mathsf{D}_{\rm{s}} h^{-2} &\left[ \left( \rho_{i,j-1} - 2 \rho_{i,j} + \rho_{i,j+1} \right) + \left( -\frac{1}{12} \rho_{i-2,j} + \frac{4}{3} \rho_{i-1,j} - \frac{5}{2} \rho_{i,j} + \frac{4}{3} \rho_{i+1,j} - \frac{1}{12} \rho_{i+2,j} \right) \right] \\
                                   &= -\frac{4 J_{y}}{3 h \mathsf{D}_{\rm{a}}} - \frac{1}{6 h^{2}} \rho_{i-2,j} + \frac{8}{3 h^{2}} \rho_{i-1,j} + \frac{6+\frac{4 \mathsf{D}_{\rm{s}}}{\mathsf{D}_{\rm{a}}}-\frac{3 \mathsf{D}_{\rm{a}}}{\mathsf{D}_{\rm{s}}}}{6 h^{2}} \rho_{i,j-1} - \frac{9}{2 h^{2}} \rho_{i,j} + \frac{6-\frac{4 \mathsf{D}_{\rm{s}}}{\mathsf{D}_{\rm{a}}}+\frac{3 \mathsf{D}_{\rm{a}}}{\mathsf{D}_{\rm{s}}}}{6 h^{2}} \rho_{i,j+1} = 0 \,.
    \end{aligned} \label{eq:left_par2perp4_2}
\end{equation}
A similar procedure can be followed for the nodes on the right, top, and bottom sides of the walls, and upon rewriting~\eqref{eq:left_par2perp4_2} we obtain for the governing equations around the walls
\begin{align}
    &\begin{aligned} \label{eq:left_par2perp4} 
        \textrm{Left:} \ &\mathsf{D}_{\rm{s}} \mathsf{D}_{\rm{a}} \rho_{i-2,j} -16 \mathsf{D}_{\rm{s}} \mathsf{D}_{\rm{a}} \rho_{i-1,j} +\left(-6 \mathsf{D}_{\rm{s}} \mathsf{D}_{\rm{a}}-4 \mathsf{D}_{\rm{s}}^2+3 \mathsf{D}_{\rm{a}}^2\right) \rho_{i,j-1} +27 \mathsf{D}_{\rm{s}} \mathsf{D}_{\rm{a}} \rho_{i,j} \\
                       &\quad +\left(-6 \mathsf{D}_{\rm{s}} \mathsf{D}_{\rm{a}}+4 \mathsf{D}_{\rm{s}}^2-3 \mathsf{D}_{\rm{a}}^2\right) \rho_{i,j+1} + 8 h \mathsf{D}_{\rm{s}} J_y = 0 \,,
	\end{aligned} \\
    &\begin{aligned} \label{eq:right_par2perp4} 
        \textrm{Right:} \ &-\left(6 \mathsf{D}_{\rm{s}} \mathsf{D}_{\rm{a}}-4 \mathsf{D}_{\rm{s}}^2+3 \mathsf{D}_{\rm{a}}^2\right) \rho_{i,j-1} + 27 \mathsf{D}_{\rm{s}} \mathsf{D}_{\rm{a}} \rho_{i,j} - 16 \mathsf{D}_{\rm{s}} \mathsf{D}_{\rm{a}} \rho_{i+1,j} + \mathsf{D}_{\rm{s}} \mathsf{D}_{\rm{a}} \rho_{i+2,j} \\
                         &\quad- \left(6 \mathsf{D}_{\rm{s}} \mathsf{D}_{\rm{a}}+4 \mathsf{D}_{\rm{s}}^2-3 \mathsf{D}_{\rm{a}}^2\right) \rho_{i,j+1} - 8 h \mathsf{D}_{\rm{s}} J_y = 0 \,,
	\end{aligned} \\
    &\begin{aligned} \label{eq:bottom_par2perp4} 
        \textrm{Bottom:} \ &-\left(6 \mathsf{D}_{\rm{s}} \mathsf{D}_{\rm{a}}-4 \mathsf{D}_{\rm{s}}^2+3 \mathsf{D}_{\rm{a}}^2\right) \rho_{i-1,j} + \mathsf{D}_{\rm{s}} \mathsf{D}_{\rm{a}} \rho_{i,j-2} - 16 \mathsf{D}_{\rm{s}} \mathsf{D}_{\rm{a}} \rho_{i,j-1} + 27 \mathsf{D}_{\rm{s}} \mathsf{D}_{\rm{a}} \rho_{i,j} \\
						 &\quad - \left(6 \mathsf{D}_{\rm{s}} \mathsf{D}_{\rm{a}}+4 \mathsf{D}_{\rm{s}}^2-3 \mathsf{D}_{\rm{a}}^2\right) \rho_{i+1,j} - 8 h \mathsf{D}_{\rm{s}} J_x = 0 \,,
	\end{aligned} \\
    &\begin{aligned} \label{eq:top_par2perp4}
        \textrm{Top:} \ &\left(-6 \mathsf{D}_{\rm{s}} \mathsf{D}_{\rm{a}}-4 \mathsf{D}_{\rm{s}}^2+3 \mathsf{D}_{\rm{a}}^2\right) \rho_{i-1,j} + 27 \mathsf{D}_{\rm{s}} \mathsf{D}_{\rm{a}} \rho_{i,j} + \left(-6 \mathsf{D}_{\rm{s}} \mathsf{D}_{\rm{a}}+4 \mathsf{D}_{\rm{s}}^2-3 \mathsf{D}_{\rm{a}}^2\right) \rho_{i+1,j} \\
                       &\quad - 16 \mathsf{D}_{\rm{s}} \mathsf{D}_{\rm{a}} \rho_{i,j+1} + \mathsf{D}_{\rm{s}} \mathsf{D}_{\rm{a}} \rho_{i,j+2} + 8 h \mathsf{D}_{\rm{s}} J_x = 0 \,.
	\end{aligned} 
\end{align}
\eqref{eq:left_par2perp4}--\eqref{eq:top_par2perp4} differ from \eqref{eq:left_par4perp2}--\eqref{eq:top_par4perp2} by the coefficients on the terms perpendicular to walls and the flux terms, leading to slight differences in the numerical results.
The equivalent of the results shown in the main text, with Fig.~\ref{fig:odd_diffusion} demonstrating the density profiles and concentration differences and ~\ref{fig:odd_diff_forces} demonstrating the forces, are shown for the fourth order and second order in the parallel and perpendicular directions scheme in Figs.~\ref{fig:odd_diffusion_2}--\ref{fig:odd_diff_forces_2}.
Figs.~\ref{fig:odd_diffusion} and \ref{fig:odd_diffusion_2} have similar concentration differences, with the values in Fig.~\ref{fig:odd_diffusion} shifted to lower values of $ \mathsf{D}_{\rm s} / \mathsf{D}_{\rm a} $ relative to Fig.~\ref{fig:odd_diffusion_2}.
The measured diffusion constants from simulation, shown in Fig.~\ref{fig:diffusion_trend}, indicate that the second and fourth order in the parallel and perpendicular directions scheme is more representative of the simulated chiral active particle systems.
Similarly, the force profile computed using the second and fourth order in the parallel and perpendicular directions scheme has a higher agreement with the simulated data than the other scheme.
We also compute concentration differences between the internal and external regions of the walls for these parameters for both schemes in Fig.~\ref{fig:odd_diff_conc}.

\begin{figure*}
    \centering
    \includegraphics[width=1.0\linewidth]{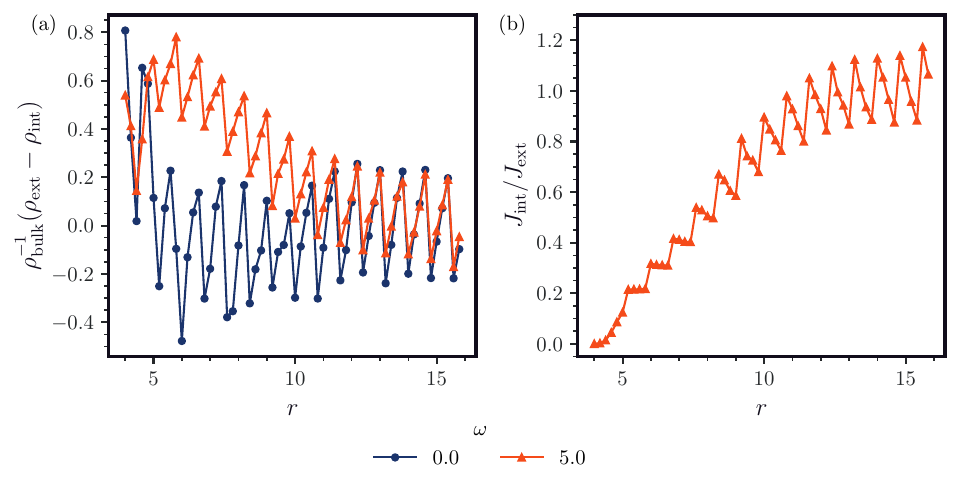}
    \caption{(a) Difference between external and internal densities and (b) ratio of internal and external fluxes near the walls for varying $ \omega $ at $ \nu = 80 $ and $ \rho = 0.2 $.}
    \label{fig:rho_diff_flux_ratio}
\end{figure*}

As mentioned, the fluxes perpendicular to the walls are zero by no-flux boundary conditions.
The fluxes parallel to the walls are measured in simulation.
To do so, the flux field with respect to the solvent is evaluated using Gaussian kernels per Sec.~\ref{sec:field_details}.
To yield the flux parallel to the walls, we average over the flux values at distances $ \frac{b}{2} ( 1 + \sin ( \frac{\pi}{3} ) ) \pm 0.4 $ perpendicularly away from the walls, where $ b $ is the hexatic lattice spacing for a given activity \cite{lauersdorf2021phase}.
The difference between the external and internal densities and the ratio of the internal and external fluxes computed using this scheme are shown in Fig.~\ref{fig:rho_diff_flux_ratio}.
These measured profiles follow a sawtooth pattern, resulting in the patterns observed in the concentration and force profiles obtained from the finite difference schemes.

\section{Methods} 

\subsection{System and integration scheme}\label{sec:appwca}

The solvent and solutes evolve under overdamped Langevin dynamics using HOOMD-blue \cite{anderson2020hoomd} for two-dimensional and three-dimensional systems.
The solvent is modeled as chiral active particles, in which the $i$th solvent particle evolves as \cite{liao2018clustering,ma2022dynamical}
\begin{align} 
    \dot{\rb}_i ( t ) & = \xi^{-1} [\Fb_i + \nu \bb_i ( t )] + \sqrt{2D_t} \Lambdab_{i},                   \\
    \bb_i ( t )    & = [\cos\theta_i ( t ), \sin\theta_i ( t )]^\top, \\
    \dot{\theta}_i ( t ) & = \omega + \sqrt{2D_r} \Gamma_{i} \,,
\end{align}
where $\rb_i$ denotes the position of the $i$th particle, $ \xi $ is the translational drag coefficient, $\Fb_i $ is the force,  $\bb_i$ denotes the direction of its active velocity, $ \nu $ is the magnitude of the active force, $ \omega $ is the active torque, and $ D_t $ and $ D_r $ are the translational and rotational diffusion constants that are related by the formula $ D_r = 3 \ \sigma^{-2} D_t $.
Here, $ \Lambdab_i $ and $ \Gamma_i $ are independent Gaussian white noises with zero mean and unit variance.
The particles interact with the Weeks-Chandler-Anderson (WCA) potential \cite{weeks1971role}, given by the sum $ U = \sum_{i\neq j} u(l_{ij}) $, where $ l_{ij} = \left| \rb_i - \rb_j \right| $ is the distance between particles $ i $ and $ j $.
The form of $ u(l_{ij}) $ is given by
\begin{equation}
    u ( l_{ij} ) = 4 \epsilon_{ij} \left[ \left( \frac{\sigma_{ij}}{l_{ij}} \right)^{12} - \left( \frac{\sigma_{ij}}{l_{ij}} \right)^{6} + \frac{1}{4} \right] \theta \left( 2^{\frac{1}{6}} - \frac{l_{ij}}{\sigma_{ij}} \right) \,, \label{eq:u_wca_22}
\end{equation}
where $ \epsilon_{ij} $ and $ \sigma_{ij} $ are the energy and length scales set by the particle types, respectively, and $ \theta $ is the Heaviside function.
The corresponding force is $ \Fb_i = - \diff{U(t)}{\rb_i} $.
The solutes are represented as rigid bodies composed of particles interacting under the WCA potential with other particles not in the same rigid body \cite{nguyen2011rigid,glaser2020pressure}.
Following Ref.~\cite{nguyen2011rigid}, a rigid body $ b $ composed of $ N_b $ internal particles has its internal particles indexed by $ B_{bk} = \left[ B_{b1}, \ldots, B_{bN_b} \right] $, with the overall rigid body position and quaternion given by $ \rb_b $ and $ \qb_b $, respectively.
Evaluating the net force and torque on the $ b $th rigid body per
\begin{align}
    \Fb_b &= \sum_{i \in B_{bk}} \Fb_i \,, \\
    \Tb_b &= \sum_{i \in B_{bk}} \left( \rb_i - \rb_b \right) \times \Fb_i \,,
\end{align}
the equations of motion for the rigid body are given by
\begin{align} 
    \dot{\rb}_b ( t ) & = \xi_b^{-1} \Fb_b + \sqrt{2D_{t,b}} \Lambdab_{b} \,, \label{eq:rigid_body_eom_t} \\
    \dot{\qb}_b ( t ) & = 0.5 (\xi_{b,r}^{-1} \Tb_b + \sqrt{2 D_{r,b}} \qb_b \Gammab_{b} \qb_b^{-1}) \qb_b \,, \label{eq:rigid_body_eom_r_2}
\end{align}
where $ \xi_b $ and $ \xi_{b,r} $ are the translational and rotational drag coefficients, respectively, $ D_{t,b} $ and $ D_{r,b} $ are the translational and rotational diffusion constants, respectively, and $ \Lambdab_b $ and $ \Gammab_b $ are independent Gaussian white noises with zero mean and unit variance.
As the system is two-dimensional, only the $ z $-component of the torque and $ \Gammab_b $ are non-zero.
The relations between the drag coefficients and diffusion constants are given by the Stokes-Einstein and Stokes-Einstein-Debye relations in which for the solute one has
\begin{align}
    D_{t,b} &= \kbT \xi_b^{-1} \,, \\
    D_{r,b} &= \kbT \xi_{b,r}^{-1} = 3 \ \sigma_{H}^{-2} D_{t,b} \,,
\end{align}
where $ \sigma_H $ is the hydrodynamic diameter of the solute, and $ k_\mathrm{B} $ and $ T $ are the Boltzmann constant and temperature, respectively.
Analogous relations hold for the solvent.
We choose the diffusion constants for the solvent and solutes to be equal with $ D_t = D_{t,b} = 1 $, $ \kbT = 1 $, and $ \epsilon_{ij} = 40 $ and $ \sigma_{ij} = 1 $ unless otherwise specified.

For simulations in three dimensions, the equations of motion for the rigid body given by Eqs.~\eqref{eq:rigid_body_eom_t} and \eqref{eq:rigid_body_eom_r_2} extend naturally, while the solvent particles evolve under slightly modified equations.
These equations of motion are given by
\begin{align} 
    \dot{\rb}_i ( t ) & = \xi^{-1} [\Fb_i + \nu \qb_i \fb_i \qb_i^{-1}] + \sqrt{2D_t} \Lambdab_{i},                   \\
    \dot{\qb}_i ( t ) & = 0.5 (\frac{\omega}{\xi_{r}} \qb_i \ub_i \qb_i^{-1} + \left(\cos \frac{\lambda_i}{2}, || \nb_i ||^{-1} \nb_i \sin \frac{\lambda_i}{2} \right) ) \qb_i, \\
    \nb_i & = \left(\qb_i \fb_i \qb_i^{-1}\right) \times \Lambdab_i \,,
\end{align}
where $ \fb_i $ is the direction of the active force in the local particle coordinate frame of the $ i $th particle, $ \ub_i $ is the direction of the active torque in the local particle coordinate frame of the $ i $th particle, $ \xi_r $ is the rotational drag coefficient, $ \lambda_i $ is Gaussian white noise with mean zero and variance $ 2 D_r $ where $ D_r $ is the rotational diffusion constant, and $ \Lambdab_i $ is a Gaussian white noise about the unit sphere.
In all cases, we set $ \fb_i = \left[ 1, 0, 0 \right] $ and $ \ub_i = \left[ 0, 0, 1 \right] $.

\subsection{Local densities} \label{sec:ld}

\begin{figure*}[h]
    \centering
    \includegraphics[width=1.0\linewidth]{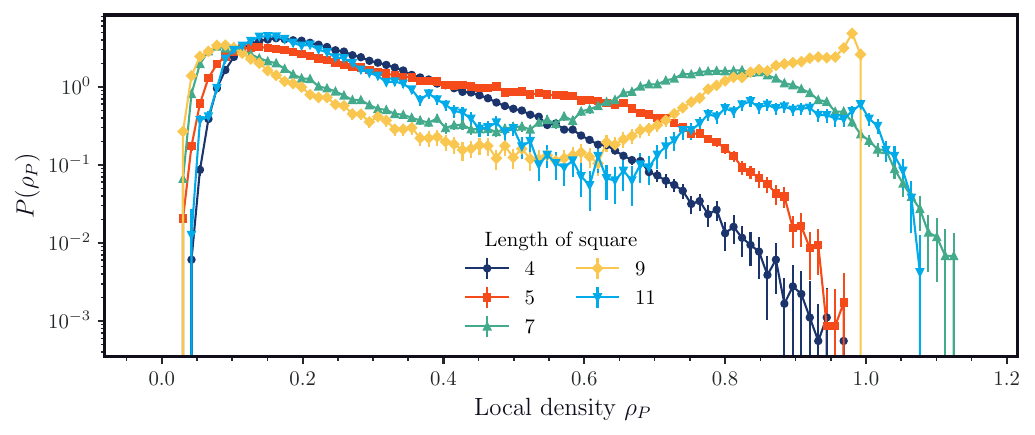}
    \caption{Histograms of the local density of passive square particles of varying length for $ \phi_{A} = 0.2 $, $ \phi_{P} = 0.2 $, $ \nu = 80 $, and $ \omega = 5 $.}
    \label{fig:voronoi_analysis_size}
\end{figure*}

\begin{figure*}[h]
    \centering
    \includegraphics[width=1.0\linewidth]{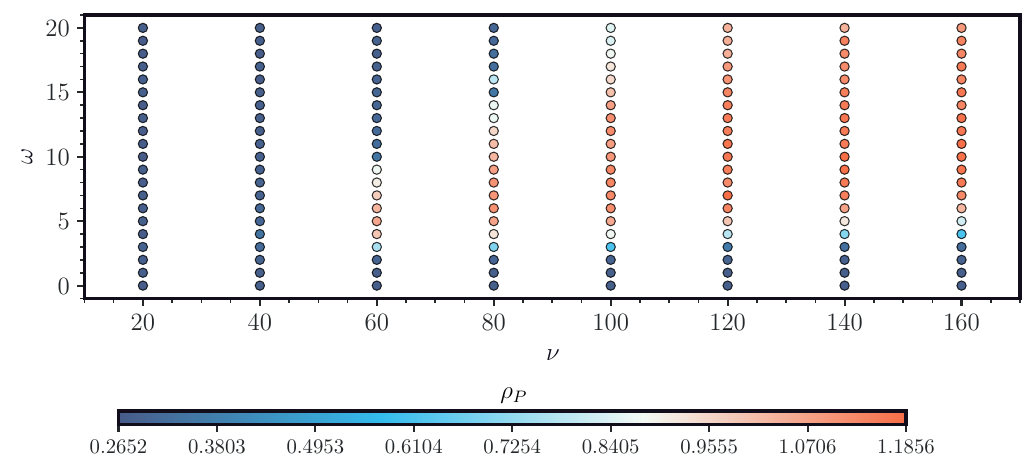}
    \caption{Average density of passive square particles in the cluster phase for varying $ \nu $ and $ \omega $ for $ \phi_{A} = 0.2 $, $ \phi_{P} = 0.2 $, and $ L = 9\ \sigma $. In cases where the system remains homogeneous, the cluster density is set equal to the overall passive particle density.}
    \label{fig:voronoi_phase_diagram}
\end{figure*}

\begin{figure*}[t]
    \centering
    \includegraphics[width=1.0\linewidth]{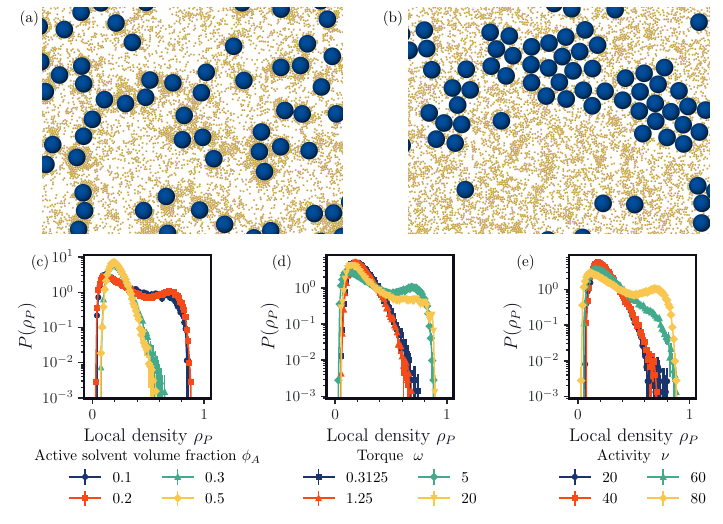}
    \caption{
    Multiple passive disks of radii $5\ \sigma$ in a chiral active Brownian particle bath with a passive volume fraction of $ 0.2 $. (a,b) Systems where the passive disks do not and do assemble for $ \phi_{A} = 0.2 $, $ \nu = 80 $, and (a) $ \omega = 0.3125 $ and (b) $ \omega = 5 $, respectively. Histograms of local density of the passive particles, $ \rho_{P} $, for (c) varying $ \phi_{A} $ at $ \omega = 5 $ and $ \nu = 80 $, (d) varying $ \omega $ at $ \phi_{A} = 0.2 $ and $ \nu = 80 $, (e) varying $ \nu $ at $ \omega = 5 $ and $ \phi_{A} = 0.2 $.
    }
    \label{fig:chiral_sa_circ}
\end{figure*}

\begin{figure*}[t]
    \centering
    \includegraphics[width=1.0\linewidth]{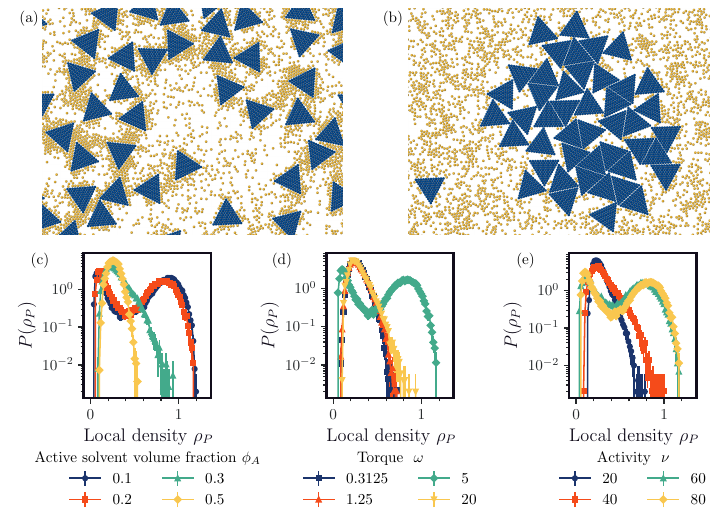}
    \caption{
    Multiple passive triangles of side length $12\ \sigma$ in a chiral active Brownian particle bath with a passive volume fraction of $ 0.2 $. (a,b) Systems where the passive triangles do not and do assemble for $ \phi_{A} = 0.2 $, $ \nu = 80 $, and (a) $ \omega = 0.3125 $ and (b) $ \omega = 5 $, respectively. Histograms of local density of the passive particles, $ \rho_{P} $, for (c) varying $ \phi_{A} $ at $ \omega = 5 $ and $ \nu = 80 $, (d) varying $ \omega $ at $ \phi_{A} = 0.2 $ and $ \nu = 80 $, (e) varying $ \nu $ at $ \omega = 5 $ and $ \phi_{A} = 0.2 $.
    }
    \label{fig:chiral_sa_tri}
\end{figure*}

\begin{figure*}[t]
    \centering
    \includegraphics[width=1.0\linewidth]{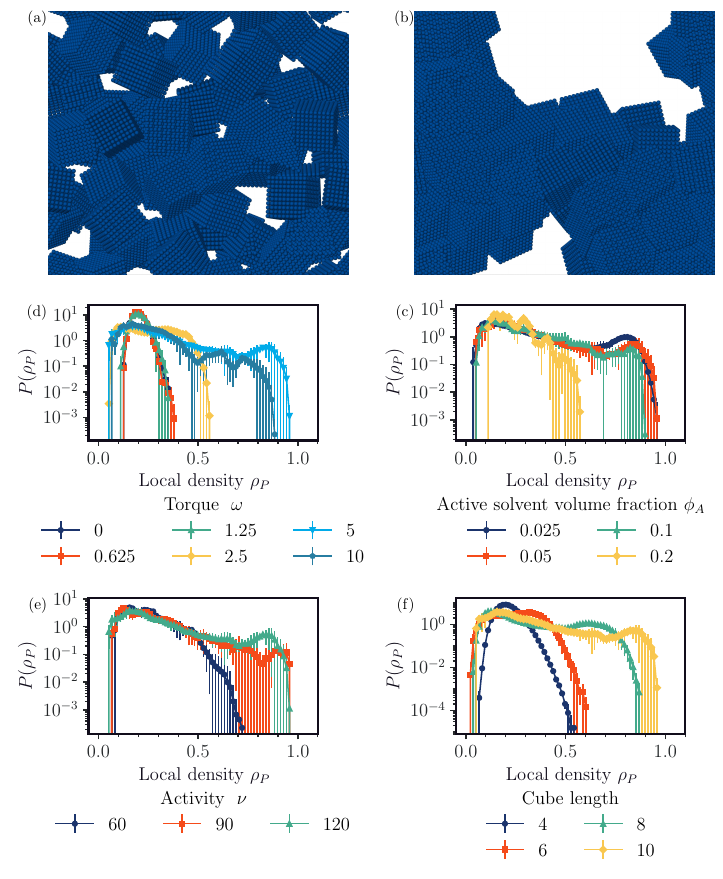}
    \caption{
        Multiple passive cubes of length $10\ \sigma $ in a chiral active Brownian particle bath with a passive volume fraction of $ 0.2 $ in three dimensions. (a,b) Systems where the passive cubes do not and do assemble for $ \phi_{A} = 0.2 $, $ \nu = 120 $, and (a) $ \omega = 0.625 $ and (b) $ \omega = 5 $, respectively, where the solvent is not visualized.
        Histograms of local density of the passive cube particles, $ \rho_{P} $, for $ \phi_{P} = 0.2 $ at (c) varying $ \omega $ with $ \phi_{A} = 0.05 $, $ \nu = 120 $ and $ L = 10\ \sigma $, (d) varying $ \phi_{A} $ at $ \nu = 120 $, $ \omega = 5 $, and $ L = 10\ \sigma $, (e) varying $ \nu $ at $ \phi_{A} = 0.05 $, $ \omega = 5 $, and $ L = 10\ \sigma $, (f) varying $ L $ at $ \phi_{A} = 0.05 $, $ \nu = 120 $, and $ \omega = 5 $.
    }
    \label{fig:chiral_sa_cube}
\end{figure*}

To obtain the local densities of passive particles, $ \rho_{P} $ (Figs.~\ref{fig:chiral_sa}, \ref{fig:chiral_sa_sphere}, and ~\ref{fig:voronoi_analysis_size}-\ref{fig:chiral_sa_cube}), a Voronoi tesselation is obtained on the positions of the passive particles via Freud \cite{ramasubramani2020freud}.
The local density of a passive rigid body particle is then given by $ \rho_{P} = N_{P} / A_{V} $, where $ N_{P} $ is the number of particles in the rigid body and $ A_{V} $ is the area of the associated Voronoi cells.
For results involving passive disk and sphere particles, the local density is instead computed per $ \rho_{P} = V_{P} / A_{V} $, where $ V_{P} $ is the volume of the particle.
In both cases this rescaling of $ \rho_{P} $ is done to keep the axes comparable between different systems and sizes of passive particles.

\subsection{Force and effective free energy} \label{sec:ffe}

The force on a fixed wall is obtained by summing over the forces from the solvent interacting with the wall.
Denoting the force on a wall as $ \Fb_{b\mathrm{p}} $, the force on the wall is given by
\begin{equation}
    \Fb_{b\mathrm{p}} = \sum_{i \in B_{b\mathrm{p}k}} \sum_{j \in N_{A}} - \diff{u ( l_{ij} )}{\lb_{ij}}  \,.
\end{equation}
Denoting the force on the left and right walls by $ \Fb_L $ and $ \Fb_R $, the effective force in the $ x $ direction is given by
\begin{equation}
    F = \frac{1}{2} \left( F_{R,x} - F_{L,x} \right) \,,
\end{equation}
which has the convention per Sec.~\ref{sec:packing} that $ F > 0 $ corresponds repulsion and $ F < 0 $ corresponds to attraction.

The effective free energy, taken as the logarithm of the radial distribution function $ - \ln g ( r ) $, is obtained by simulating two passive walls.
A tether potential of the form \cite{noguchi2005dynamics}
\begin{equation}
    U_{\mathrm{tether}} ( r ) = 
    \begin{cases}
        k_{\mathrm{tether}} \frac{\exp ( 1 / (l_{0}-r))}{l_{\mathrm{max}}-r}, &\text{ if } r > l_{0} \\
        0, &\text{ otherwise } 
    \end{cases}
\end{equation}
is used to constrain the walls below a cutoff distance $ l_{\mathrm{max}} $. 
The radial distribution function is then obtained in the standard manner by binning the distances between the walls obtained in simulation and normalizing relative to the ideal gas distribution \cite{allen2017computer}.

\subsection{Gaussian kernel} \label{sec:field_details}

A Gaussian kernel is used to convert per-particle quantities into field quantities using Freud \cite{ramasubramani2020freud}.
The Gaussian kernel is given by, for a quantity $ k $,
\begin{equation}
    k ( \xb ) = C^{-1} \sum_{i \in \mathrm{Nb}} k_i\exp \left( - \frac{(\xb-\rb_i)^{2}}{2 \sigma^{2}} \right) \,, \label{eq:gaussian_kernel}
\end{equation}
where the summation denotes the particles about $ \xb $ within a cutoff distance $ r_{\mathrm{max}} $.
The normalization constant $ C $ in 2D is evaluated as
\begin{equation}
    C = \int_{0}^{r_{\mathrm{max}}} 2 \pi r \exp \left( - \frac{r^{2}}{2 \sigma^{2}} \right) = 2 \pi \sigma^{2} \left( 1 - \exp ( -\frac{r_{\mathrm{max}}^{2}}{2 \sigma^{2}} ) \right) \,.
\end{equation}
We take $ r_{\mathrm{max}} = 0.5 $ and $ \sigma = 1 $ to yield statistics corresponding to local regions corresponding to single particles.
The kernel is evaluated on a grid cell with $ \frac{\sigma}{15} $ spacing.

This procedure is used to evaluate the density, flux, and orientation fields.
The density field $ \rho ( \xb ) $ is evaluated by setting $ k_i = 1 $.
The flux field is evaluated per $ \kb_i = \Fb_i + \nu \bb_i $.
For the orientation field, as the angle $ \theta $ is a periodic variable, $ \theta ( \xb ) $ is evaluated through a circular mean.
This is done by evaluating $ \cos \theta ( \xb ) $ and $ \sin \theta ( \xb ) $ through $ k_i = \cos \theta_i $ and $ k_i = \sin \theta_i $, respectively, and then obtain the orientation field as
\begin{equation}
    \theta ( \xb ) = \atantwo \left( \sin \theta ( \xb ), \cos \theta ( \xb ) \right) \,.
\end{equation}

\subsection{Density and hexatic order parameter between walls} \label{sec:dh}

The local density is obtained by enumerating the number of particles between the walls and dividing by the free area between the walls.
The hexatic order parameter of a particle is obtained through Freud \cite{ramasubramani2020freud} by evaluating~\eqref{eq:hexatic}.
In the case of there being no particles between the walls, the hexatic order parameter is taken to be zero.

\subsection{Constraining the system} \label{sec:cs}

To perform simulations of moving walls with fixed orientation and $ y $-positions, the orientation is fixed by setting the rotational drag coefficient to be numerically infinite, and the $ y $-position is reset to the initial $ y $-position after each time step.

\subsection{Glass model} \label{sec:glass}

We modify a model used in the simulation of glass formers due to the model's resistance to crystallization \cite{paoluzzi2022motility,ninarello2017models}.
In our model, values of $ \sigma_i $ are drawn from a discrete probability distribution $ P ( \sigma ) = A \sigma^{-3} $ with $ 51 $ bins between $ \sigma_{\mathrm{min}} = 0.6 $ and $ \sigma_{\mathrm{max}} = 2.29 $, with the value of $ \sigma_{\mathrm{min}} $ set to a predetermined value and $ \sigma_{\mathrm{max}} $ set to allow the particles to have the same volume fraction as a monodisperse system with $ \sigma = 1 $ and $ \rho = 0.4 $.
The value of $ A $ is set to ensure that the distribution is normalized.
The value of $ \epsilon_{ij} $ is set to $ 40 $, and the value of $ \sigma_{ij} $ is set to the arithmetic mean of $ \sigma_i $ and $ \sigma_j $.
The WCA potential of~\eqref{eq:u_wca_22} is used to model the interaction between particles.
This model defers from the original glass model of Refs.~\cite{paoluzzi2022motility,ninarello2017models} in that the form of the potential is different along with the mixing rule for $ \sigma_{ij} $, along with the polydispersity being drawn from a discrete distribution rather than a continuous one.

\section{Other supplementary data}

\begin{figure*}[h]
    \centering
    \includegraphics[width=1.0\linewidth]{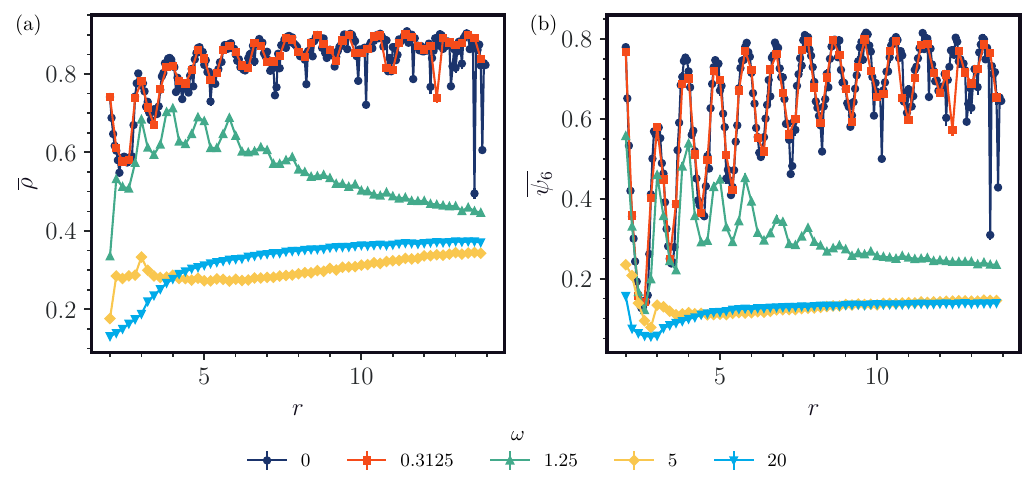}
    \caption{(a) Average density $ \overline{\rho} $ and (b) hexatic order parameter $ \overline{\psi}_6 $ between two passive walls of length $ 10 $, bulk $ \rho = 0.4 $, and $ \nu = 80 $ for varying torque and separation length. See Sec.~\ref{sec:dh} for further details on how the density and hexatic order parameter are computed.}
    \label{fig:wall_rho_hex_torque_80}
\end{figure*}

\begin{figure*}[h]
    \centering
    \includegraphics[width=1.0\linewidth]{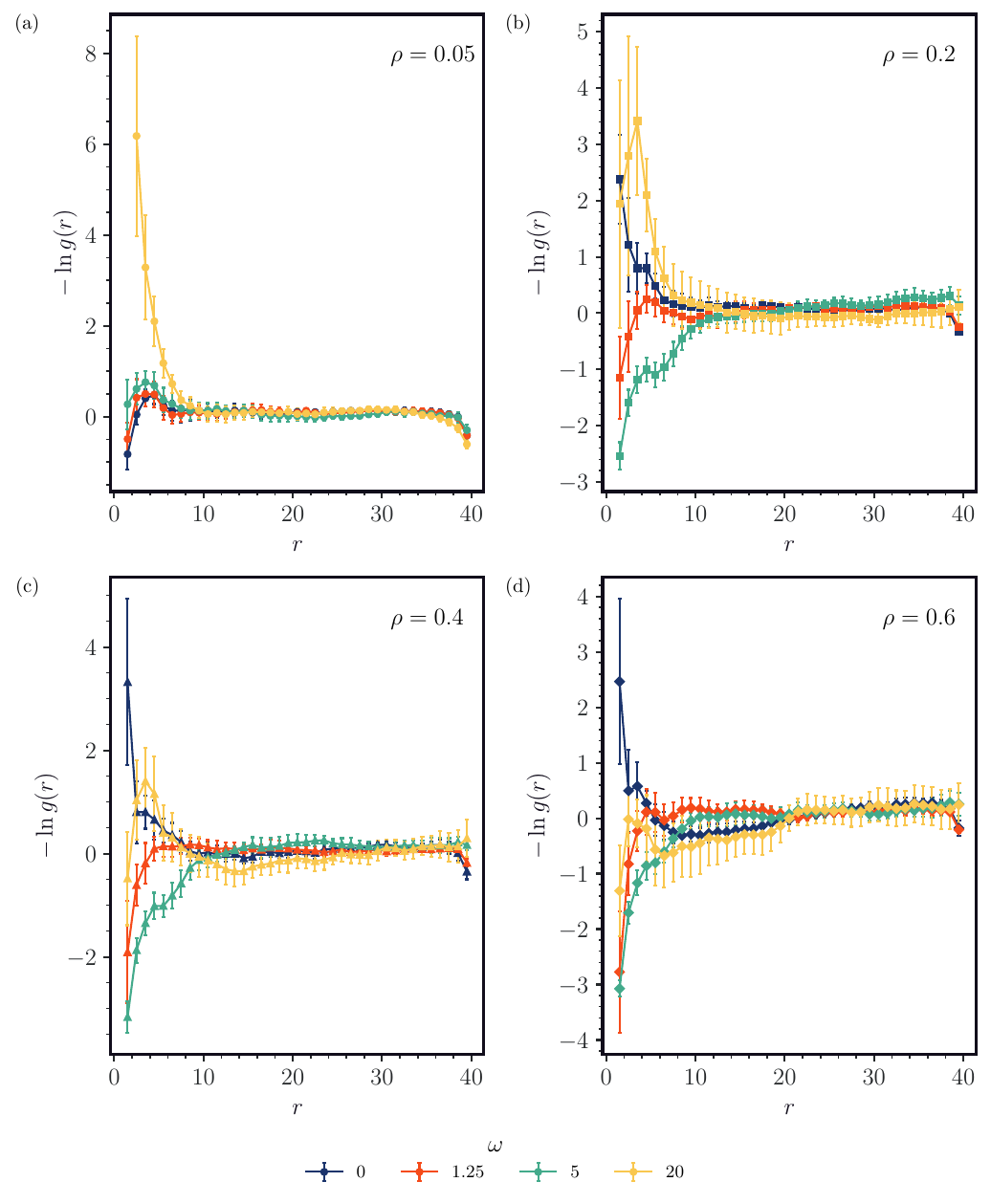}
    \caption{Effective free energy between two passive walls of length $ 10 $ and $ \nu = 80 $ for varying $ \rho $ and $ \omega $.}
    \label{fig:free_energy_density}
\end{figure*}

\begin{figure*}
    \centering
    \includegraphics[width=1.0\linewidth]{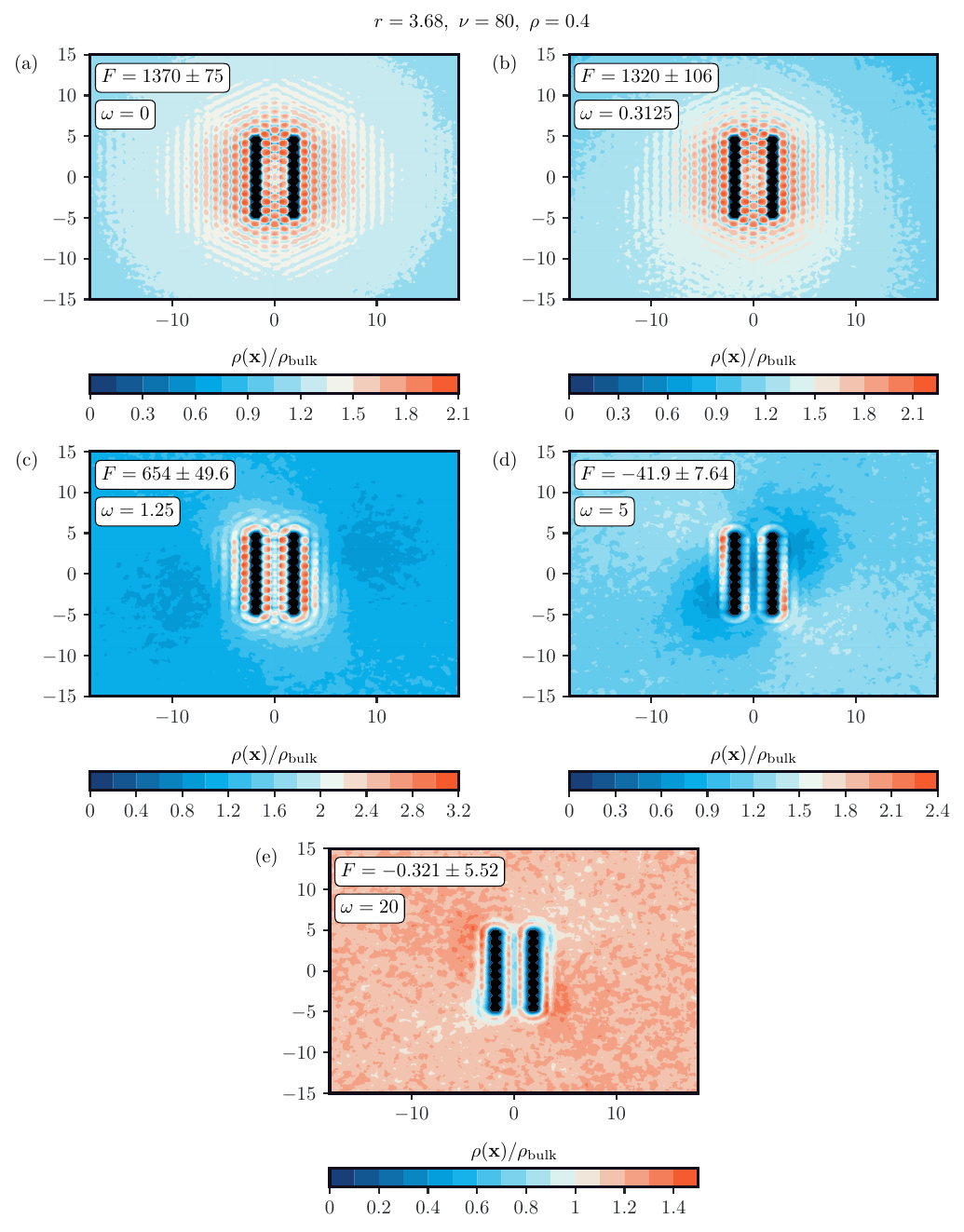}
    \caption{Density fields for maximal repulsive force between two passive walls in a chiral active Brownian particle bath.}
    \label{fig:gd_0}
\end{figure*}

\begin{figure*}
    \centering
    \includegraphics[width=1.0\linewidth]{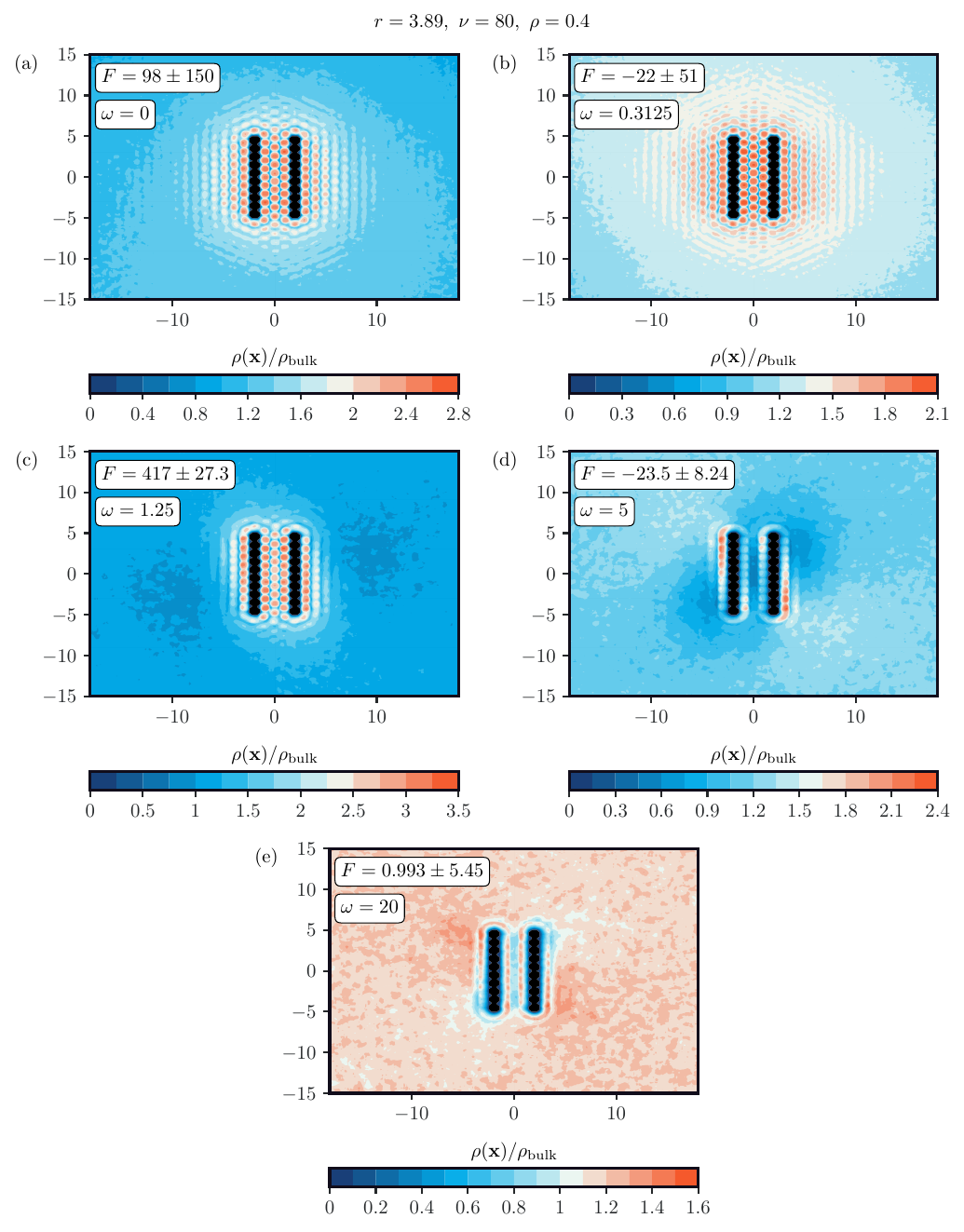}
    \caption{Density fields for a separation with near zero force between two passive walls in a chiral active Brownian particle bath.}
    \label{fig:gd_1}
\end{figure*}

\begin{figure*}
    \centering
    \includegraphics[width=1.0\linewidth]{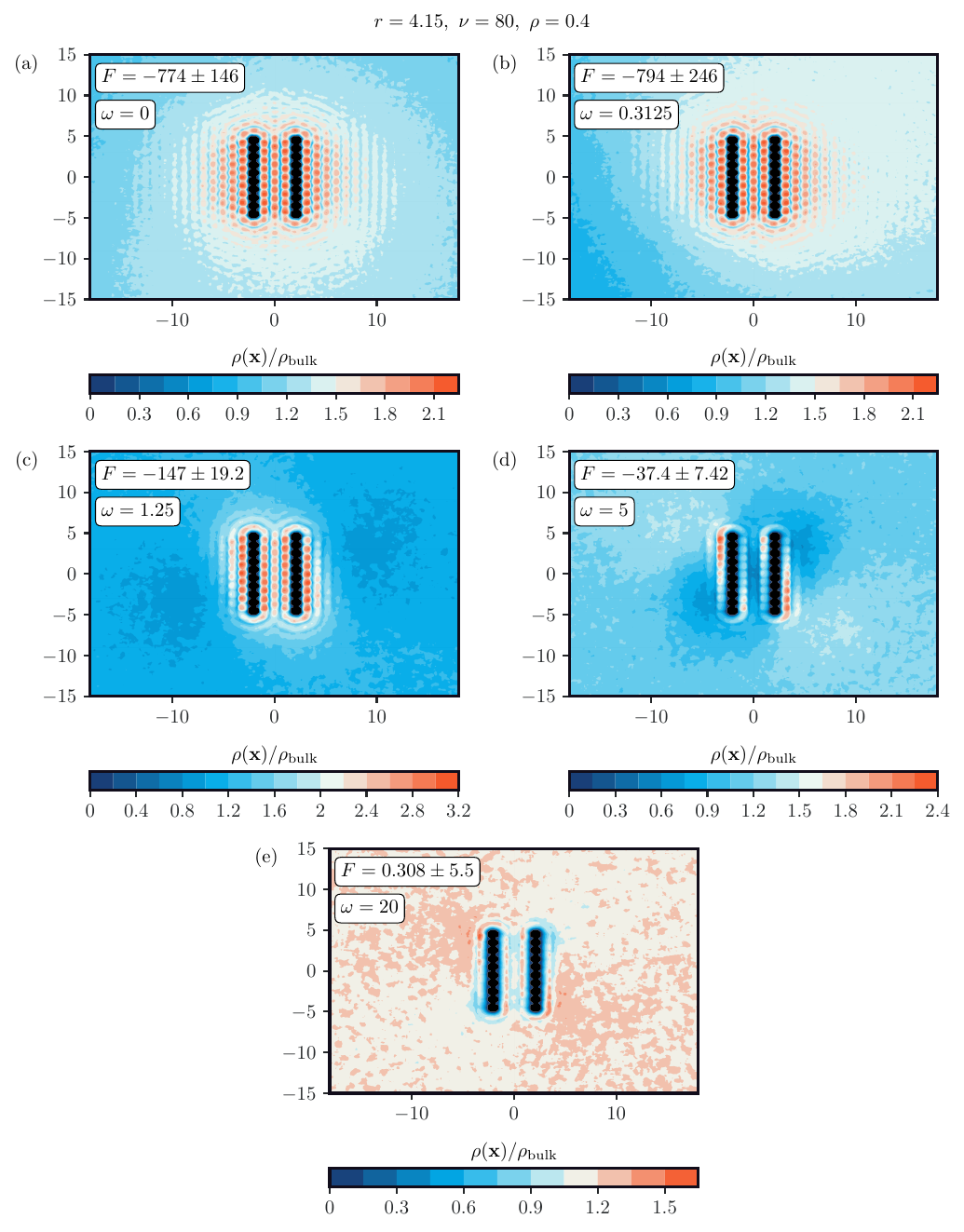}
    \caption{Density fields for maximal attractive force between two passive walls in a chiral active Brownian particle bath.}
    \label{fig:gd_2}
\end{figure*}

\begin{figure*}
    \centering
    \includegraphics[width=1.0\linewidth]{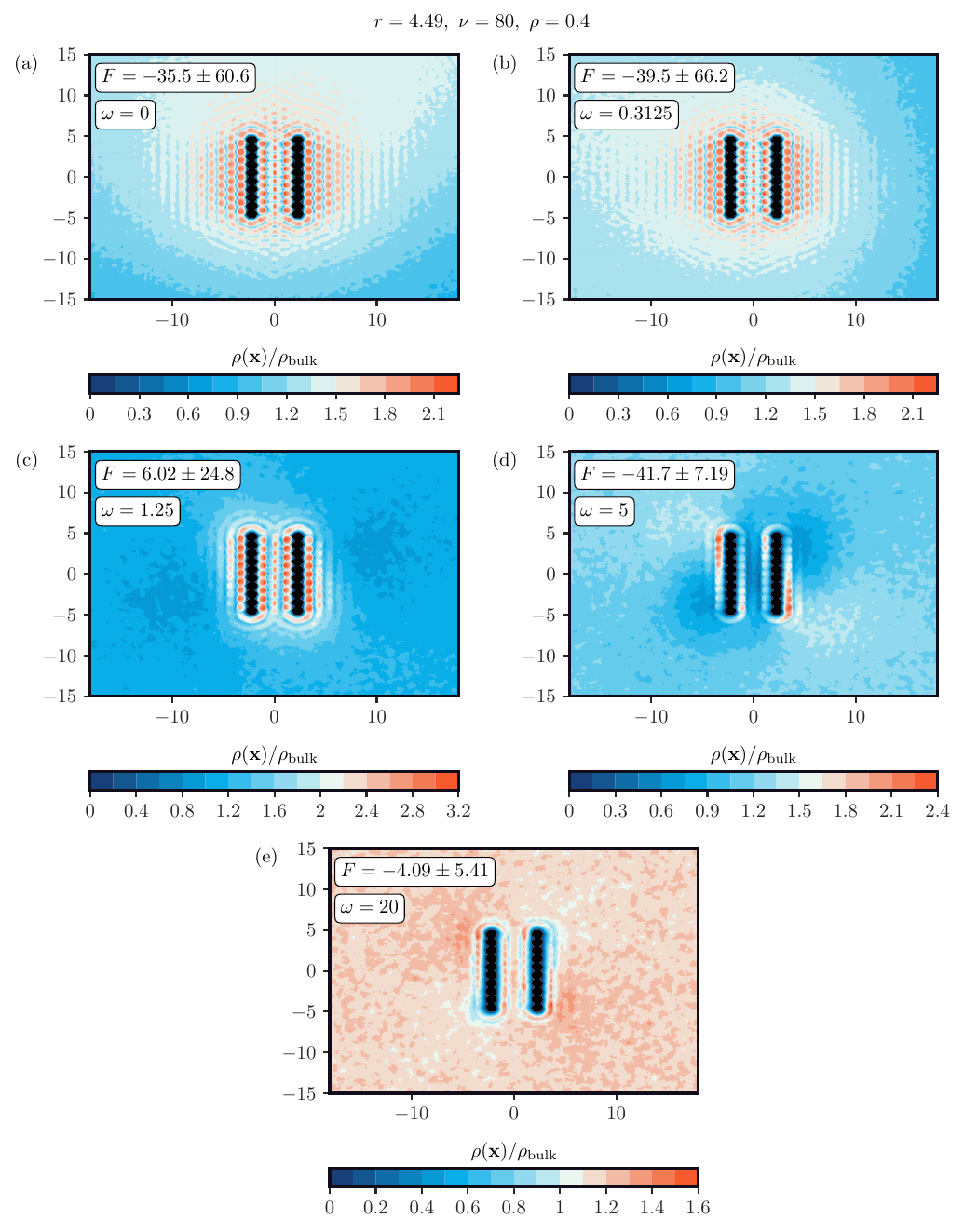}
    \caption{Density fields for a separation with near zero force between two passive walls in a chiral active Brownian particle bath.}
    \label{fig:gd_3}
\end{figure*}

\begin{figure*}
    \centering
    \includegraphics[width=1.0\linewidth]{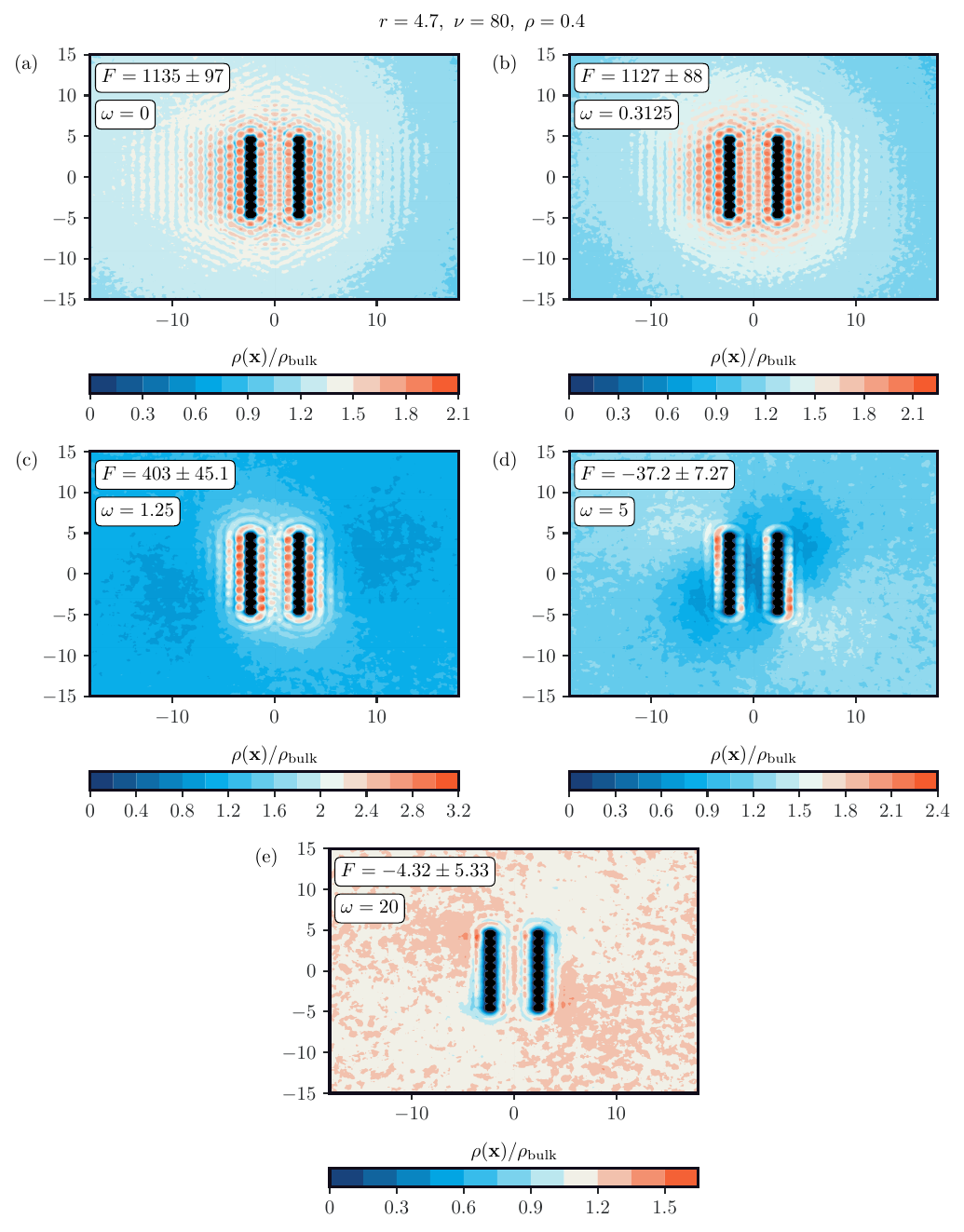}
    \caption{Density fields for maximal repulsive force between two passive walls in a chiral active Brownian particle bath.}
    \label{fig:gd_4}
\end{figure*}

\begin{figure*}
    \centering
    \includegraphics[width=1.0\linewidth]{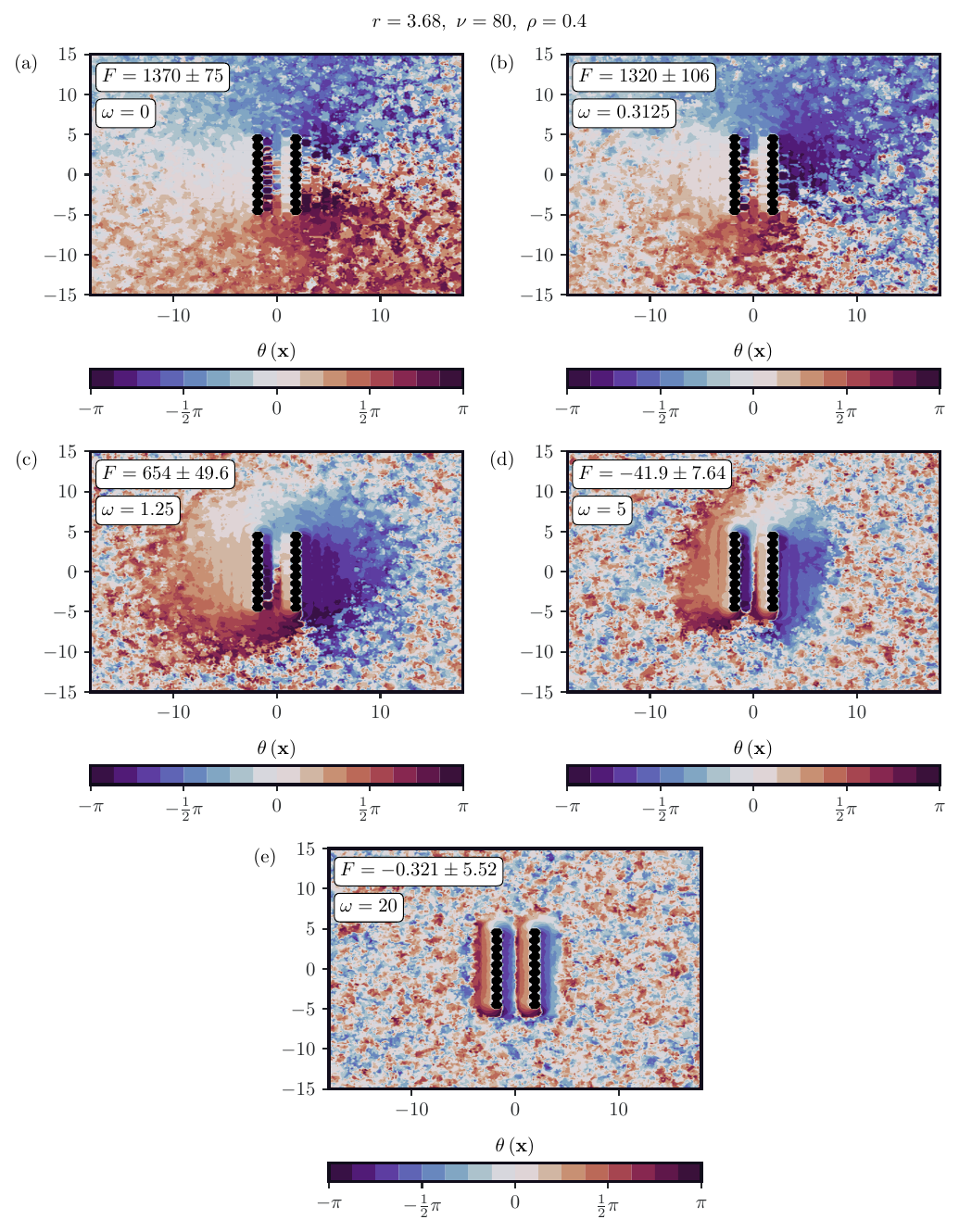}
    \caption{Orientation fields for maximal repulsive force between two passive walls in a chiral active Brownian particle bath.}
    \label{fig:go_0}
\end{figure*}

\begin{figure*}
    \centering
    \includegraphics[width=1.0\linewidth]{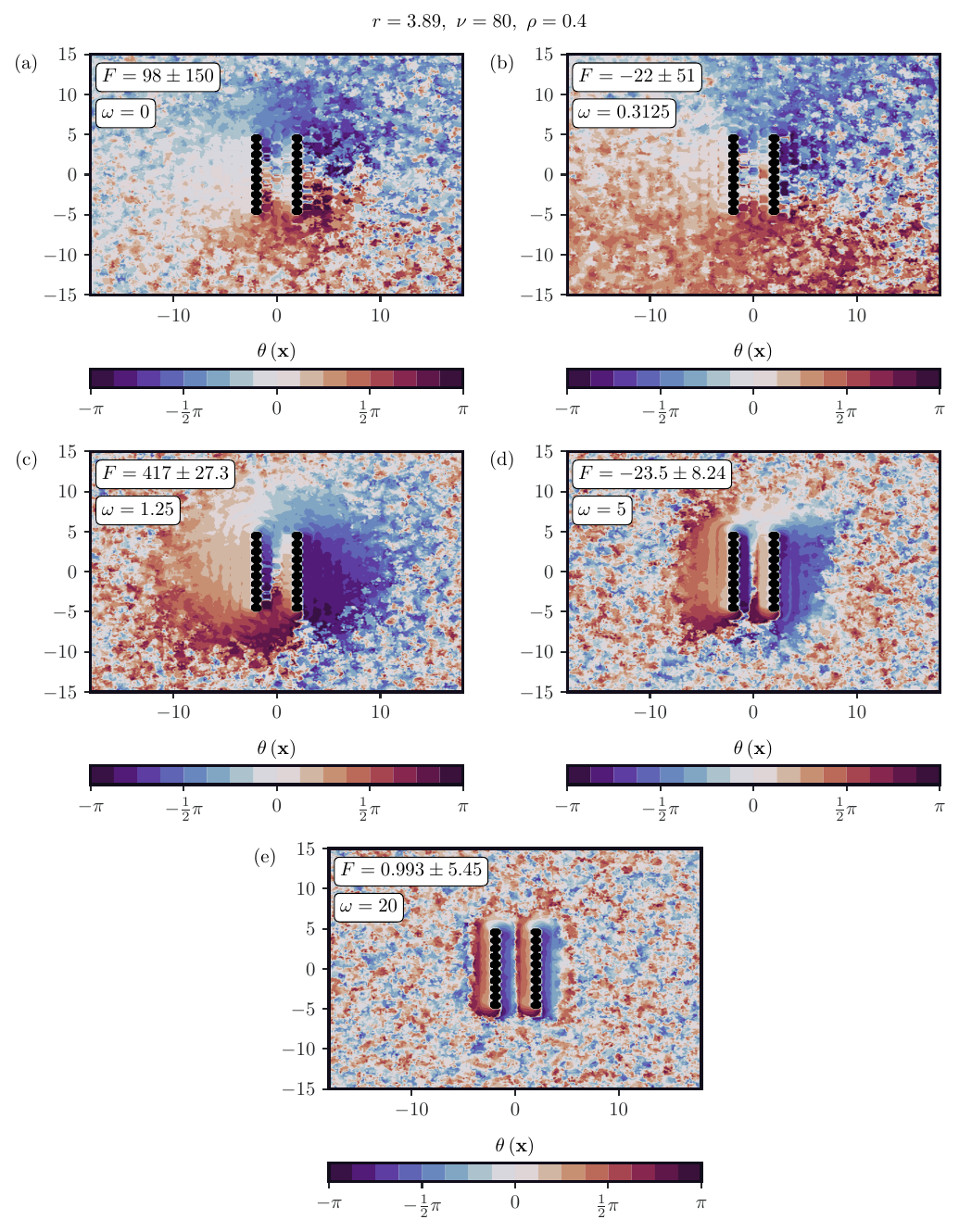}
    \caption{Orientation fields for a separation with near zero force between two passive walls in a chiral active Brownian particle bath.}
    \label{fig:go_1}
\end{figure*}

\begin{figure*}
    \centering
    \includegraphics[width=1.0\linewidth]{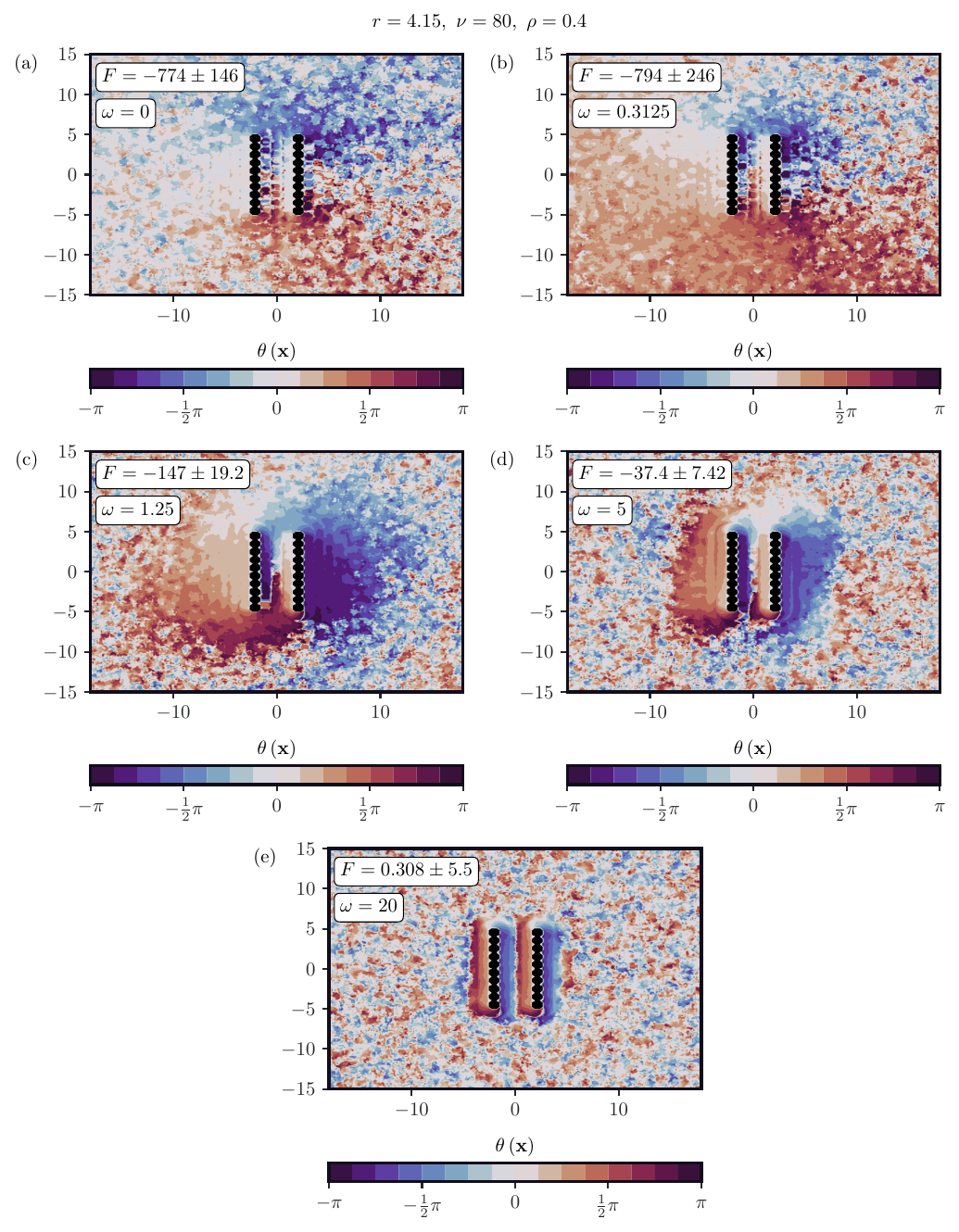}
    \caption{Orientation fields for maximal attractive force between two passive walls in a chiral active Brownian particle bath.}
    \label{fig:go_2}
\end{figure*}

\begin{figure*}
    \centering
    \includegraphics[width=1.0\linewidth]{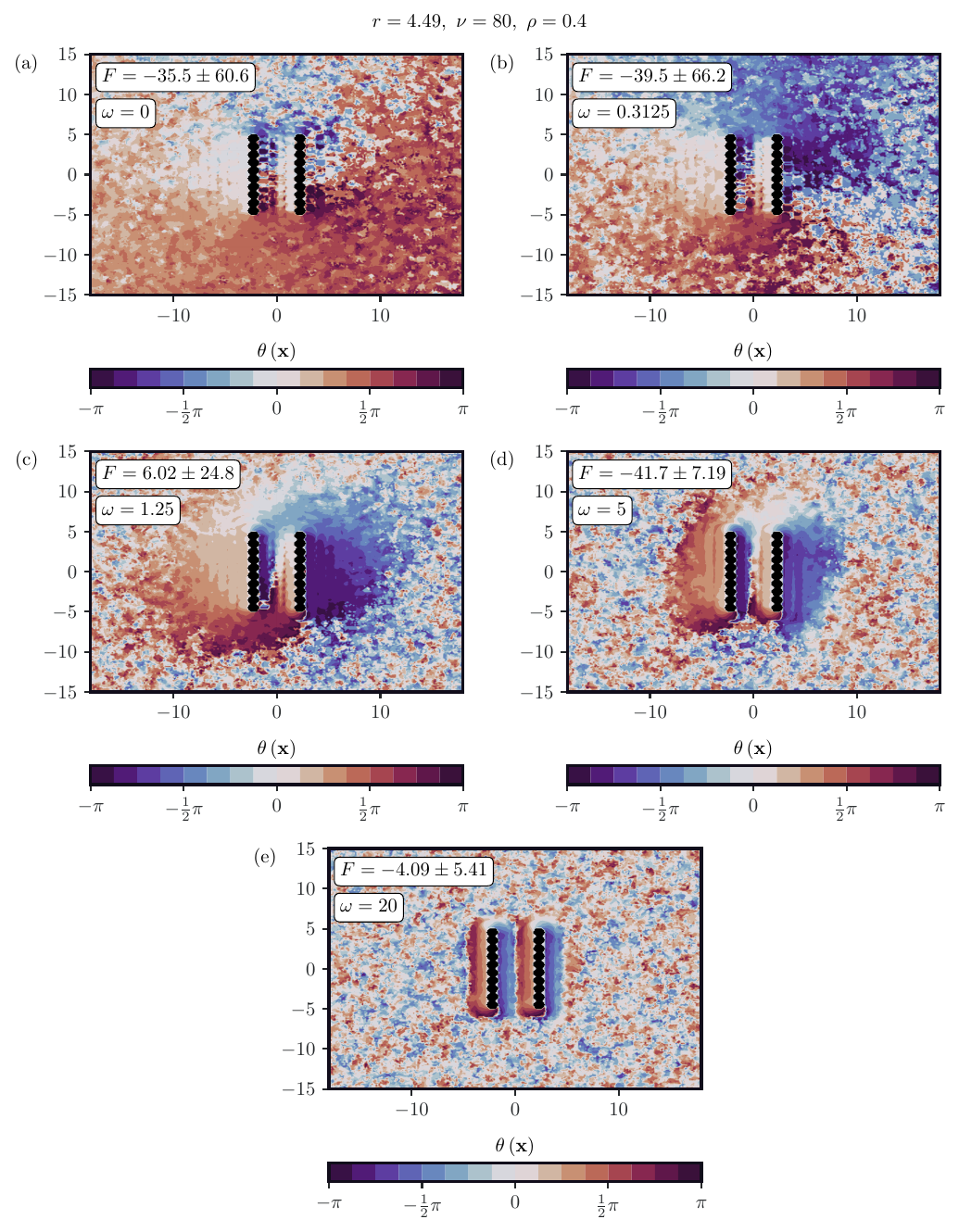}
    \caption{Orientation fields for a separation with near zero force between two passive walls in a chiral active Brownian particle bath.}
    \label{fig:go_3}
\end{figure*}

\begin{figure*}
    \centering
    \includegraphics[width=1.0\linewidth]{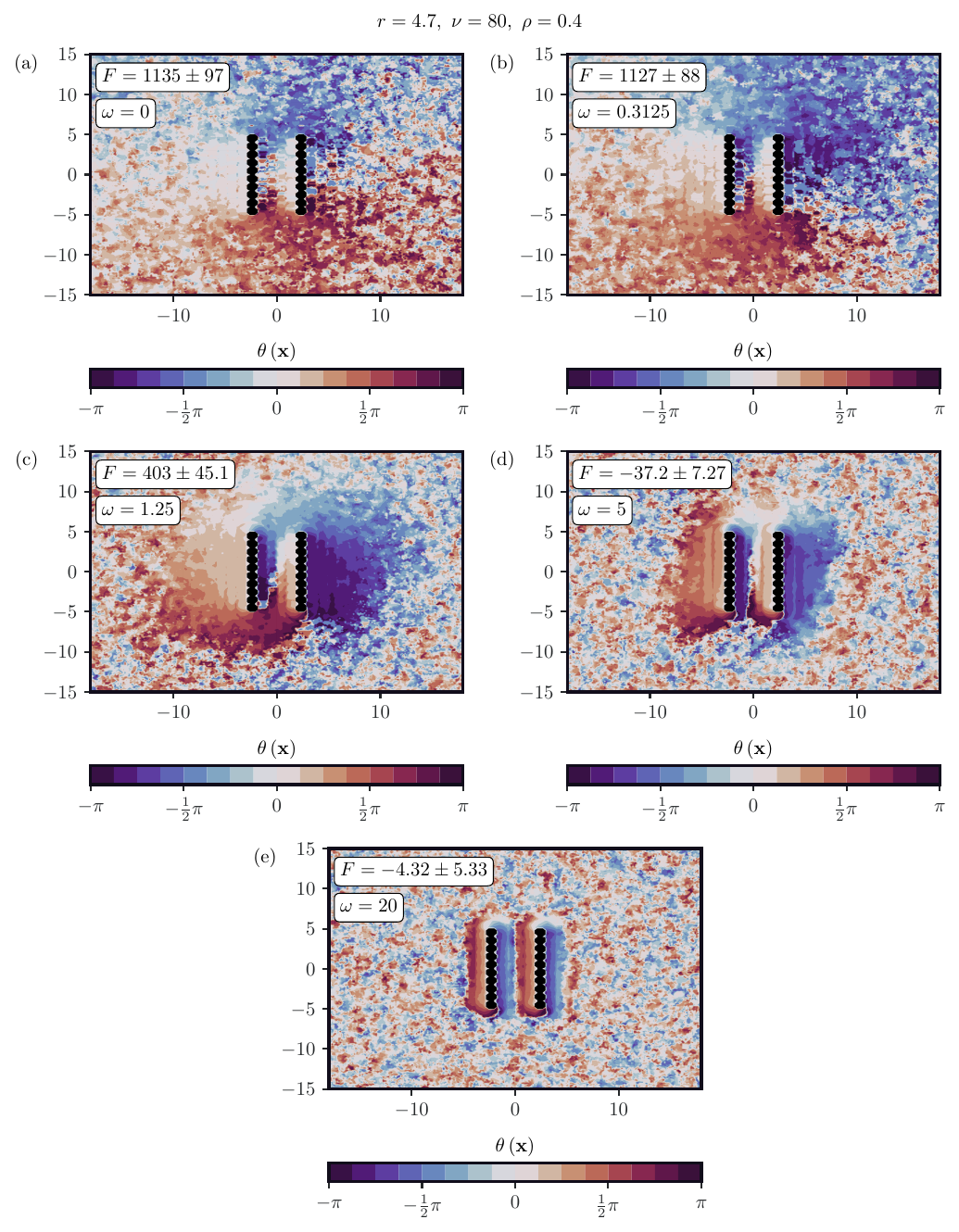}
    \caption{Orientation fields for maximal repulsive force between two passive walls in a chiral active Brownian particle bath.}
    \label{fig:go_4}
\end{figure*}

\end{document}